\documentclass[12pt,a4paper]{article}
\pdfoutput=1
\usepackage{jheppub}
\usepackage{feynmp}
\usepackage{caption}
\usepackage{subcaption}
\usepackage{shuffle}
\usepackage{amssymb}
\usepackage{amsmath}
\usepackage[utf8]{inputenc}

\def\usegraph#1#2{\includegraphics[scale=1.0,trim=0 #1 0 0]{graphs/#2.pdf}}

\def\braket#1{\langle #1 \rangle}

\def\Res_#1{\operatorname*{Res}_{#1}}

\def\d{\mathrm{d}}

\def\SpDenom5{\braket{12}\braket{23}\braket{34}\braket{45}\braket{51}}

\title{Overcoming Obstacles to Colour-Kinematics Duality at Two Loops}

\author[a]{Gustav Mogull,}
\author[a,b]{Donal O'Connell}

\affiliation[a]{Higgs Centre for Theoretical Physics, School of Physics and Astronomy,\\ The University of Edinburgh, Edinburgh EH9 3JZ, Scotland, UK}
\affiliation[b]{Kavli Institute for Theoretical Physics,  \\University of California, Santa Barbara, CA 93106-4030, USA}

\abstract{
The discovery of colour-kinematics duality has allowed great progress in our understanding of the UV structure of gravity. However, it has proven difficult to find numerators which satisfy colour-kinematics duality in certain cases. We discuss obstacles to building a set of such numerators in the context of the five-gluon amplitude with all helicities positive at two loops. We are able to overcome the obstacles by adding more loop momentum to our numerator to accommodate tension between the values of certain cuts and the symmetries of certain diagrams. At the same time, we maintain control over the size of our ansatz by identifying a highly constraining but desirable symmetry property of our master numerator. The resulting numerators have twelve powers of loop momenta rather than the seven one would expect from the Feynman rules.
}
\preprint{Edinburgh 2015/29}

\begin{document}
\maketitle

\section{Introduction}\label{sec:intro}

The last decade or so has seen rapid progress in our understanding of scattering amplitudes in quantum field theories.
Several excellent reviews of the field are now available; see, for example~\cite{Roiban:2010kk,Ellis:2011cr, Bern:2011qt, Carrasco:2011hw, Dixon:2013uaa, Elvang:2013cua, Carrasco:2015iwa}. This paper is concerned with one of these recent insights, due to Bern, Carrasco and Johansson (BCJ): colour-kinematics duality and the associated double copy relation between gauge theory and gravity~\cite{Bern:2008qj, Bern:2010ue, Bern:2010yg}.

Colour-kinematics duality is the statement that numerators of trivalent Feynman-like diagrams may be chosen such that they satisfy the same algebraic relations as the colour factors associated with the same diagrams (namely, the numerators satisfy the same antisymmetry and Jacobi properties as the colour factors.) At tree level, this fact leads to a set of identities known as BCJ identities~\cite{Bern:2008qj} which have been proven using a variety of techniques~\cite{BjerrumBohr:2009rd,Stieberger:2009hq,Feng:2010my,BjerrumBohr:2012mg,Cachazo:2012uq}. The existence of the numerators themselves has also been proven~\cite{Mafra:2011kj,BjerrumBohr:2012mg}. The double copy relation states that gravitational scattering amplitudes can be deduced from gauge scattering amplitudes, expressed in a form where colour-kinematics duality holds, by simply replacing the colour factors with a second set of gauge theoretic kinematic numerators. At tree level, this is equivalent to the celebrated KLT relations~\cite{Kawai:1985xq, Bern:2010yg}. 

At loop level, however, the situation is less clear. The existence of numerators which satisfy the requirements of colour-kinematics duality (for brevity, we will call these colour-dual numerators in what follows) remains a conjecture in general. Several infinite families of numerators exist at one loop; for example, in the case of pure Yang-Mills scattering amplitudes with all helicities equal, or only one different helicity~\cite{Boels:2013bi} and, more recently, for MHV scattering amplitudes at one loop in maximal $\mathcal{N}=4$ supersymmetric Yang-Mills (SYM)~\cite{He:2015wgf}.

A major reason for our interest in colour-dual numerators is that they provide a promising route towards understanding the ultraviolet structure of supergravity.  This has motivated calculations of colour-dual numerators at four points in $\mathcal{N}=4$ SYM, which are now available at up to four loops~\cite{Bern:2012uf}. There is some flexibility in the structure of the double copy---although the double copy requires two sets of gauge theory numerators, they may be from different gauge theories and only one set of numerators needs to be colour-dual. Therefore the availability of a selection of colour-dual numerators in gauge theory has also allowed rapid progress in our understanding of non-maximally supersymmetric gravity~\cite{Bern:2011rj, BoucherVeronneau:2011qv, Bern:2012cd, Bern:2012gh, Carrasco:2012ca, Bern:2013yya, Nohle:2013bfa, Bern:2013qca, Bern:2013uka, Carrasco:2013ypa, Bern:2014lha, Bern:2014sna, Bern:2015xsa}. The flexibility of the double-copy allows the construction of a range of interesting different theories of gravity; understanding the structure of this set of gravity theories has developed into a vigorous area of research~\cite{Broedel:2012rc, Chiodaroli:2013upa, Chiodaroli:2014xia, Johansson:2014zca, Johansson:2015oia, Chiodaroli:2015rdg}.

A particularly insightful approach to the double copy originates in string theory. Of course, the KLT relations have their origin in string theory, but more recently the pure spinor approach to string theory~\cite{Berkovits:2000fe} has led to a number of important results about the structure and existence of colour-dual numerators~\cite{Mafra:2011kj,Mafra:2014oia,Mafra:2014gja,Mafra:2015mja}.
Various other extensions of colour-kinematics duality and the double copy are available: to three dimensions, where the existence of Chern-Simons matter theories in addition to Yang-Mills theory leads to an intriguing web of relationships~\cite{Bargheer:2012gv, Huang:2012wr, Huang:2013kca}; to form factors~\cite{Boels:2012ew}; more general numerator representations~\cite{Bern:2011ia}; and to classical field backgrounds~\cite{Monteiro:2011pc, Saotome:2012vy, Monteiro:2014cda, Luna:2015paa}.

There is intense interest in colour-kinematics duality at more than four loops~\cite{Bern:2011qn}. However, construction of a set of numerators for the five-loop, four-point, $\mathcal{N} = 4$ SYM amplitude has proven to be difficult. An expression for the integrand of the amplitude is known~\cite{Bern:2012uc}, but finding an equivalent set of colour-dual numerators has been problematic. Given the large scale of the problem, it has also been difficult to locate the precise nature of the obstruction. This has motivated interest in finding scattering amplitudes which are simple enough to understand, but complicated enough that the colour-dual numerators are elusive. The idea is to understand the nature of obstructions to the existence of colour-dual numerators, with a view to identifying methods for overcoming these obstructions. For example, one recent suggestion is that the requirement of colour-kinematics duality can be relaxed so that they only hold on unitarity cuts~\cite{Bern:2015ooa}.

The topic of this paper is the five-point, two-loop amplitude in pure Yang-Mills theory with all helicities equal. The integrand of this amplitude is known~\cite{Badger:2013gxa,Badger:2015lda}, and, indeed, the (remarkably simple) integrated planar amplitude was recently determined~\cite{Gehrmann:2015bfy}. Moreover, the structure of the amplitude is known to be closely related to the five-point, two-loop amplitude in maximally supersymmetric Yang-Mills theory, which was constructed in colour-dual form by Carrasco and Johansson~\cite{Carrasco:2011mn}. As we will see, the problem of computing colour-dual numerators for the all-plus amplitude is surprisingly complicated. We find an obstruction to the existence of a set of colour-dual numerators containing at most 7 powers of loop momenta (as one would expect from power counting the Feynman rules.) This obstruction can be described as a tension between the value of one cut, and the symmetry properties of one of our graphs.

We resolve this tension by introducing extra powers of loop momentum into our numerators, obtaining in the end a set of numerators with 12 powers of loop momentum. It is typically a dangerous idea to consider such high powers of loop momenta, for the practical reason that a general ansatz with such high power counting will contain many terms. We circumvent this problem by identifying a desirable symmetry property of our BCJ master numerator. This symmetry is highly constraining, which made it quite feasible for us to increase the amount of loop momenta in our numerators.

The rest of our paper is organised as follows. We open with a review of the known colour-dual numerators of the two-loop, all-plus, four-point scattering amplitude~\cite{Bern:2013yya}. As we will discuss, these numerators have a very suggestive structure which we find generalises to the five point case. We then describe our construction of a set of five-point colour-dual numerators, including a discussion of the obstructions, before we conclude.

\section{The Four-Point, Two-Loop BCJ System}\label{sec:4points}

As an instructive warm up to the five-point, two-loop calculation,
we begin with a discussion of the recent calculation~\cite{Bern:2013yya} of a set of BCJ
numerators for the four-point, two-loop, all-plus case. As we will see, some
aspects of the five-point system are closely analogous to the four-point case.

\subsection{BCJ master numerators}

BCJ numerators are interrelated through a set of kinematic Jacobi identities.
Many of these identities can be used to define a subset of graph numerators in terms of
a small set of numerators known as master numerators.\footnote{The
	set of masters is not unique, but given a set of BCJ master numerators, the numerators of all other graphs in the problem are uniquely determined.}
For example, at four points the double-triangle graph is a difference of double-box graphs:
\begin{equation}\label{eq:n330example}
n\bigg(\usegraph{9}{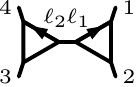}\bigg)
=n\bigg(\usegraph{9}{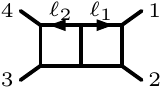}\bigg)-n\bigg(\usegraph{9}{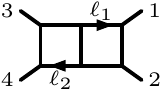}\bigg).
\end{equation}
Thus, if the double-box diagrams are masters, the double triangle is determined.

In the case of the two-loop, four-point amplitude in pure Yang-Mills theory, one can chose a set of two masters. Appropriate numerators of these master graphs were found to be~\cite{Bern:2013yya}:
\begin{subequations}\label{eq:4ptmastersold}
\begin{align}
\label{eq:n331old}
n\bigg(\usegraph{9}{delta331l}\bigg)
&=s\,F_1(\mu_1,\mu_2)+\frac{1}{2}(D_s-2)^2\mu_{11}\mu_{22}(\ell_1+\ell_2)^2
+(\ell_1+\ell_2)^2F_2(\mu_1,\mu_2),\\
\label{eq:n322old}
n\bigg(\!\usegraph{9}{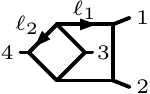}\bigg)
&=s\,F_1(\mu_1,\mu_2),
\end{align}
\end{subequations}
where $s=(p_1+p_2)^2$ is a Mandelstam invariant and $F_1(\mu_1, \mu_2)$ and $F_2(\mu_1, \mu_2)$ are functions of extra-dimensional (regulator) components $\mu_1$ and $\mu_2$ of the loop momenta. That is, working in dimensional regulation with $d= 4 -2 \epsilon$, the loop momenta $\ell_i$ consist of a four dimensional part $\bar{\ell}_i$ and a $(-2 \epsilon)$-dimensional part $\mu_i$: $\ell_i = (\bar{\ell}_i,\mu_i)$. Rotational invariance in these extra dimensions forces the $\mu_i$ in $F_1$ and $F_2$ to appear in the combinations $\mu_{ij}\equiv\mu_i \cdot\mu_j$. Specifically, the functions are defined by
\begin{subequations}\label{eq:extradimfns}
\begin{align}
\label{eq:F1}
F_1(\mu_{1},\mu_{2})
&=(D_s-2)(\mu_{11}\mu_{22}+\mu_{11}\mu_{33}+\mu_{22}\mu_{33})+16(\mu_{12}^2-\mu_{11}\mu_{22}),\\
\label{eq:F2}
F_2(\mu_{1},\mu_{2})&=F_1(\mu_1,\mu_2)-F_1(\mu_1,-\mu_2)\nonumber\\
&=4(D_s-2)\mu_{12}(\mu_{11}+\mu_{22}),
\end{align}
\end{subequations}
with $\mu_{33}=(\mu_1+\mu_2)^2=\mu_{11}+\mu_{22}+2\mu_{12}$.
The spin dimension of the gluon, $D_s$,
is $4$ in the four-dimensional helicity scheme
and $4-2\epsilon$ in conventional dimensional regulation~\cite{Bern:2002zk}.
We have extracted a factor of
\begin{align}\label{eq:4ptprefactor}
\mathcal{T}=\frac{\left[12\right]\!\left[34\right]}{\left<12\right>\!\left<34\right>},
\end{align}
which multiplies every numerator; this is convenient in view of the permutation invariance of $\mathcal{T}$ and the fact that it does not depend on loop momentum. Of course $\mathcal{T}$ is physically important: it carries the little group weight of the amplitude.

All other numerators in this system can be computed from these two masters
using appropriate Jacobi identities (a complete list was provided in ref.~\cite{Bern:2013yya}). Therefore we call the rest of the numerators descendants of the masters. Since the set of Jacobi relations are simple linear combinations of graphs in various different orderings (perhaps with shifted loop momenta), the descendent numerators are all local functions of external and loop momenta since the master numerators have this property.

Some of the descendent numerators are associated with bubble or tadpole graphs. These graphs are potentially troublesome, because intermediate propagators can be ill-defined. However, all such diagrams in this case contain scaleless integrals which vanish upon integration using dimensional regulation. Notice that all bubble-on-external-leg graphs contain two powers of $\mu$, which ensures that the diagrams are well-defined in the limit $d \rightarrow 4$ from above. A useful discussion of these subtleties is contained in~\cite{Nohle:2013bfa}. We therefore ignore these graphs.

We will determine a set of BCJ master numerators for the five-point, two-loop, all-plus Yang-Mills amplitude below. In structure, our numerators will be very similar to the numerators given in eq.~\eqref{eq:n331old}. The similarity can be made even closer by observing that the final term in the double box, eq. (\ref{eq:n331old}), namely $(\ell_1+\ell_2)^2F_2$, is not necessary.
To see this, notice that the set of Jacobi equations is a set of linear equations. Consider setting the double box to this term alone and setting the nonplanar master to zero --- that is, take
\begin{align}
\label{eq:fakemasters}
n\bigg(\usegraph{9}{delta331l}\bigg)
=(\ell_1+\ell_2)^2F_2(\mu_{1},\mu_{2}),\qquad
n\bigg(\!\usegraph{9}{delta322l}\bigg)
=0.
\end{align}
We will now show that the resulting amplitude contribution vanishes
and that all descendent symmetries and automorphisms are satisfied. 
Linearity of the system then allows us to conclude that we may
omit the term $(\ell_1+\ell_2)^2F_2$ from eq. \eqref{eq:n331old}.

Starting from the masters in eq. \eqref{eq:fakemasters}, the only diagram that gives a nonzero contribution
upon integration besides the double box is the double triangle,
determined by the Jacobi identity in eq. (\ref{eq:n330example}) to be
\begin{align}\label{eq:n330}
n\bigg(\usegraph{9}{n330l}\bigg)
&=n\bigg(\usegraph{9}{delta331l}\bigg)-n\bigg(\usegraph{9}{delta331l1243}\bigg) \\
&= (\ell_1^2 + \ell_2^2 + (\ell_1 - p_{12})^2 + (\ell_2 - p_{34})^2 -s) F_2(\mu_{1}, \mu_{2}).
\end{align}
Thus, the complete contribution to the full colour-dressed amplitude is
\begin{align}\label{eq:4ptnullcontributio}
\mathcal{A}&=ig^6\mathcal{T}\sum_{\sigma\in S_4}\sigma\circ
\bigg\{\frac{1}{4}c\bigg(\usegraph{9}{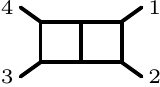}\bigg)I\bigg(\usegraph{9}{delta331l}\bigg)\left[(\ell_1+\ell_2)^2F_2\right]\nonumber\\
&\qquad +\frac{1}{8}c\bigg(\usegraph{9}{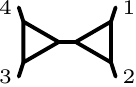}\bigg)I\bigg(\usegraph{9}{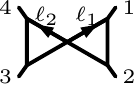}\bigg)\left[s^{-1}(\ell_1^2+\ell_2^2+(\ell_1-p_{12})^2+(\ell_2+p_{12})^2-s)F_2\right]\bigg\}\nonumber \\
&=ig^6\mathcal{T}\sum_{\sigma\in S_4}\sigma\circ
\bigg\{\frac{1}{4}c\bigg(\usegraph{9}{delta331i}\bigg)I\bigg(\usegraph{9}{delta330l}\bigg)\left[F_2\right]
\nonumber\\
&\qquad
-\frac{1}{8}c\bigg(\usegraph{9}{n330i}\bigg)I\bigg(\usegraph{9}{delta330l}\bigg)\left[F_2\right]\bigg\},
\end{align}
where we cancelled propagators in numerators and denominators and disposed of scaleless integrals. 
The integration operator for a given diagram acts on its argument as
\begin{align}
I_i\left[\mathcal{P}(p_i,\ell_i,\mu_i)\right]
\equiv\int\frac{d^d\ell_1d^d\ell_2}{(2\pi)^{2d}}\frac{\mathcal{P}(p_i,\ell_i,\mu_i)}{D_i}.
\end{align}
Since the integrands in the two terms are the same, we may combine them and use the Jacobi identity (\ref{eq:n330}) on the colour factors to find
\begin{align}
\mathcal{A}&=
ig^6\mathcal{T}\sum_{\sigma\in S_4}\sigma\circ\left\{\frac{1}{8}\left(c\bigg(\usegraph{9}{delta331i}\bigg)+c\bigg(\usegraph{9}{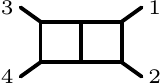}\bigg)\right)I\bigg(\usegraph{9}{delta330l}\bigg)\left[F_2(\mu_1, \mu_2)\right]\right\}\nonumber\\
&=ig^6\mathcal{T}\sum_{\sigma\in S_4}\sigma\circ\left\{\frac{1}{8}c\bigg(\usegraph{9}{delta331i}\bigg)I\bigg(\usegraph{9}{delta330l}\bigg)\left[F_2(\mu_1,\mu_2)+F_2(\mu_1,-\mu_2)\right]\right\} \nonumber\\
&=0,
\end{align}
where we exploited the sum on $S_4$ permutations to relabel $p_3\leftrightarrow p_4$
while also shifting $\ell_2\to-\ell_2-p_{12}$ in the second integral.
Finally, we have used the fact that $F_2(\mu_1,\mu_2)=-F_2(\mu_1,-\mu_2)$.

It remains to check that symmetries and automorphisms of all descendent graphs are satisfied: we have done this exhaustively for all graphs; while this is not difficult since many are vanishing, we also found private code written by Tristan Dennen to be very helpful~\cite{tristanCode}.  Consequently, an equivalent set of master numerators is
\begin{subequations}\label{eq:4ptmasters}
\begin{align}
\label{eq:n331}
n\bigg(\usegraph{9}{delta331l}\bigg)
&=s\,F_1(\mu_1, \mu_2)+\frac{1}{2}(D_s-2)^2\mu_{11}\mu_{22}(\ell_1+\ell_2)^2,\\
\label{eq:n322}
n\bigg(\!\usegraph{9}{delta322l}\bigg)
&=s\,F_1(\mu_1, \mu_2).
\end{align}
\end{subequations}
We will find a closely related set of five-point two-loop masters below.

\subsection{The maximally supersymmetric subsector}

A set of BCJ master numerators is also available~\cite{Bern:2008qj} for the two-loop, $\mathcal{N} =4$ four-point amplitude. They are simply given by
\begin{subequations}\label{eq:4ptN4numerators}
\begin{align}
n^{[\mathcal{N}=4]}\bigg(\usegraph{9}{delta331i}\bigg) &= s \, \delta^8(Q), \\
n^{[\mathcal{N}=4]}\bigg(\!\usegraph{9}{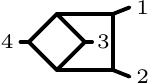}\bigg) &=s \, \delta^8(Q),
\end{align}
\end{subequations}
where $\delta^8(Q)$ is a supersymmetric delta function.
Comparing these master numerators to the simplified all-plus master numerators given in eqs. (\ref{eq:4ptmasters}) reveals an $\mathcal{N}=4$ SYM subsector of the all-plus amplitude,
generated by terms containing up to one power of $(D_s-2)$.
The function of extra-dimensional components given in eq. (\ref{eq:F1}), $F_1$,
plays the role of the supersymmetric delta function $\delta^8(Q)$.
However, while $\delta^8(Q)$ is invariant under all shifts of external and loop momenta,
$F_1$ does not quite share this property, so this subsector is not closed under Jacobi relations.

In any given Jacobi relation, all propagators but one are the same among the three diagrams. We can
therefore associate each relation with one propagator. 
For our purposes, it us useful to divide the set of Jacobi relations into two categories:
those that preserve the $(\ell_1+\ell_2)^2$ propagator in (\ref{eq:n331}),
and those that act on it.
It is easy to see that Jacobi moves of the first kind leave $F_1$ unaffected,
so descendent diagrams which can be formed using only this category of Jacobi relations belong to the $\mathcal{N}=4$ subsector.
An example of the second kind of move was given in eq. (\ref{eq:n330example}),
where if we evaluate the numerator we find
\begin{align}\label{eq:n3302}
n\bigg(\usegraph{9}{n330l}\bigg)
=s\,F_2(\mu_1,\mu_2)+\frac{1}{2}(D_s-2)^2\mu_{11}\mu_{22}\left((\ell_1+\ell_2)^2-(\ell_1-\ell_2-p_{12})^2\right),
\end{align}
using $F_2(\mu_1, \mu_2) = F_1(\mu_1, \mu_2) - F_1(\mu_1, -\mu_2)$.
This numerator vanishes in the supersymmetric case as it contains triangles;
the fact that $F_1(\mu_1, \mu_2)$ does not have the full symmetries of $\delta^8(Q)$ now prevents this from happening.
We shall refer to diagrams of this kind as ``butterflies'':
the particular BCJ moves that generate them from the masters take them outside the $\mathcal{N}=4$ subsector.

\subsection{Spanning cuts}

\begin{figure}[t]
\centering
\begin{subfigure}{0.25\textwidth}
  \centering
  \includegraphics[width=0.8\textwidth]{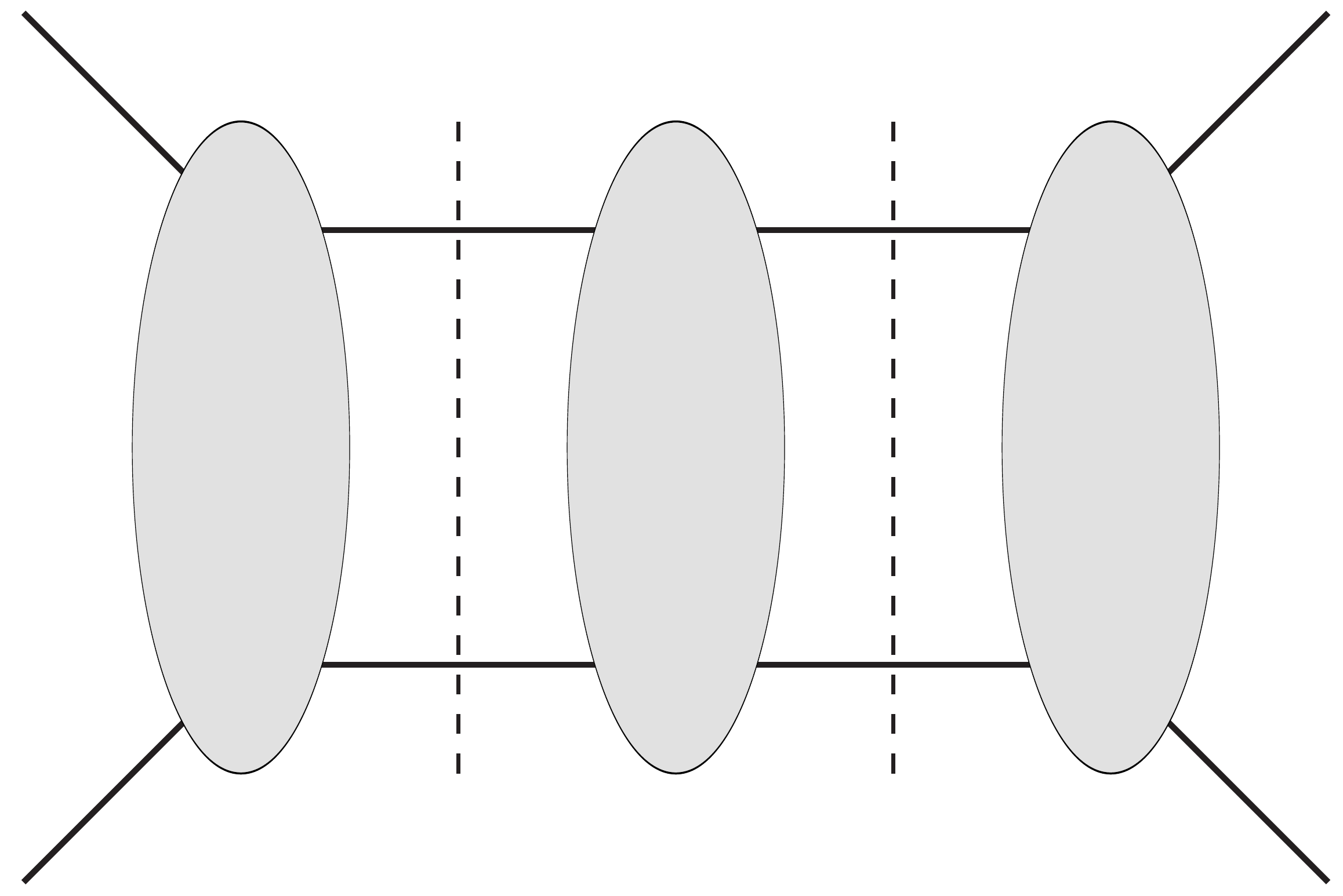}
  \caption[a]{}
\end{subfigure}
\begin{subfigure}{0.25\textwidth}
  \centering
  \includegraphics[width=0.8\textwidth]{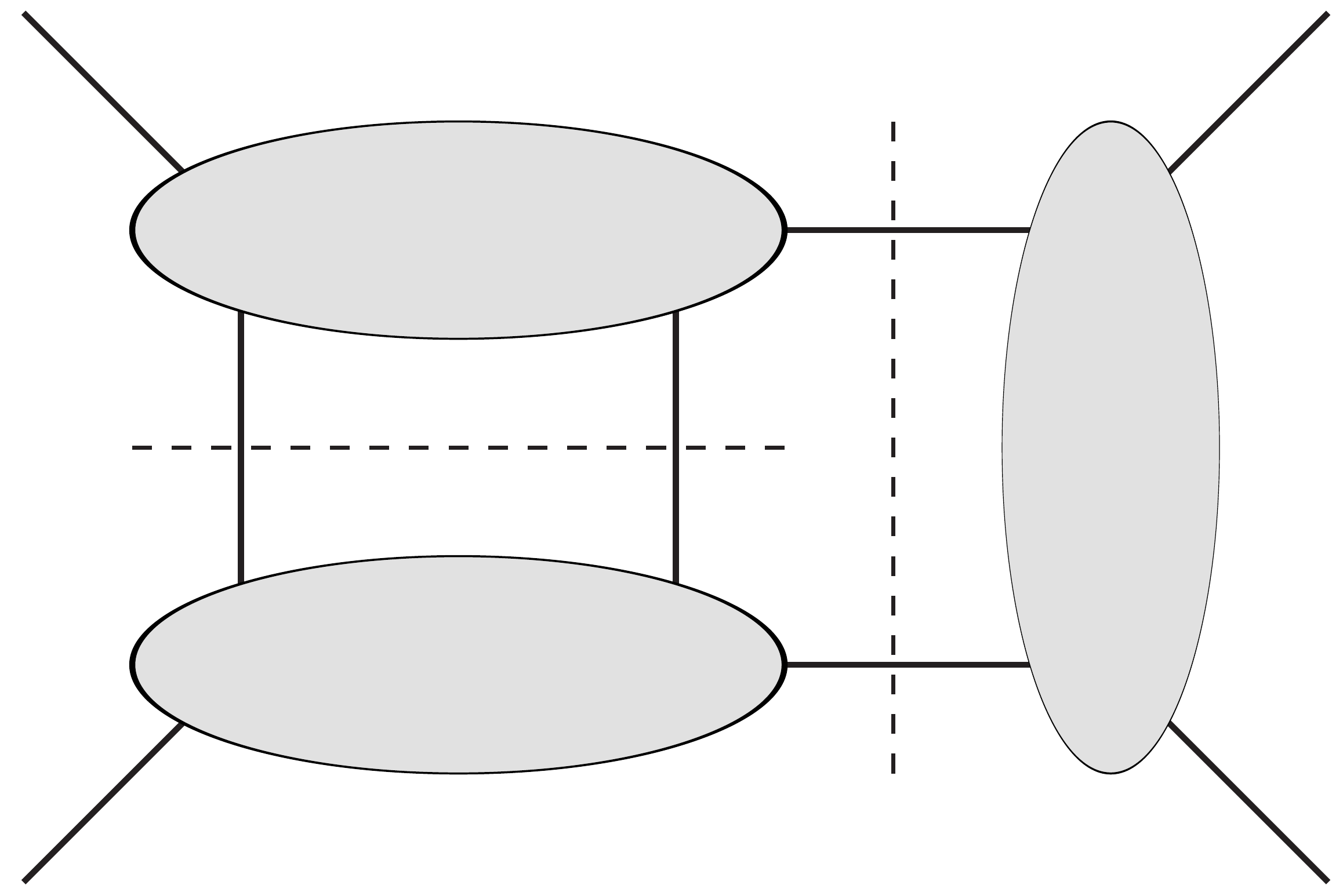}
  \caption[b]{}
\end{subfigure}
\begin{subfigure}{0.25\textwidth}
  \centering
  \includegraphics[width=0.8\textwidth]{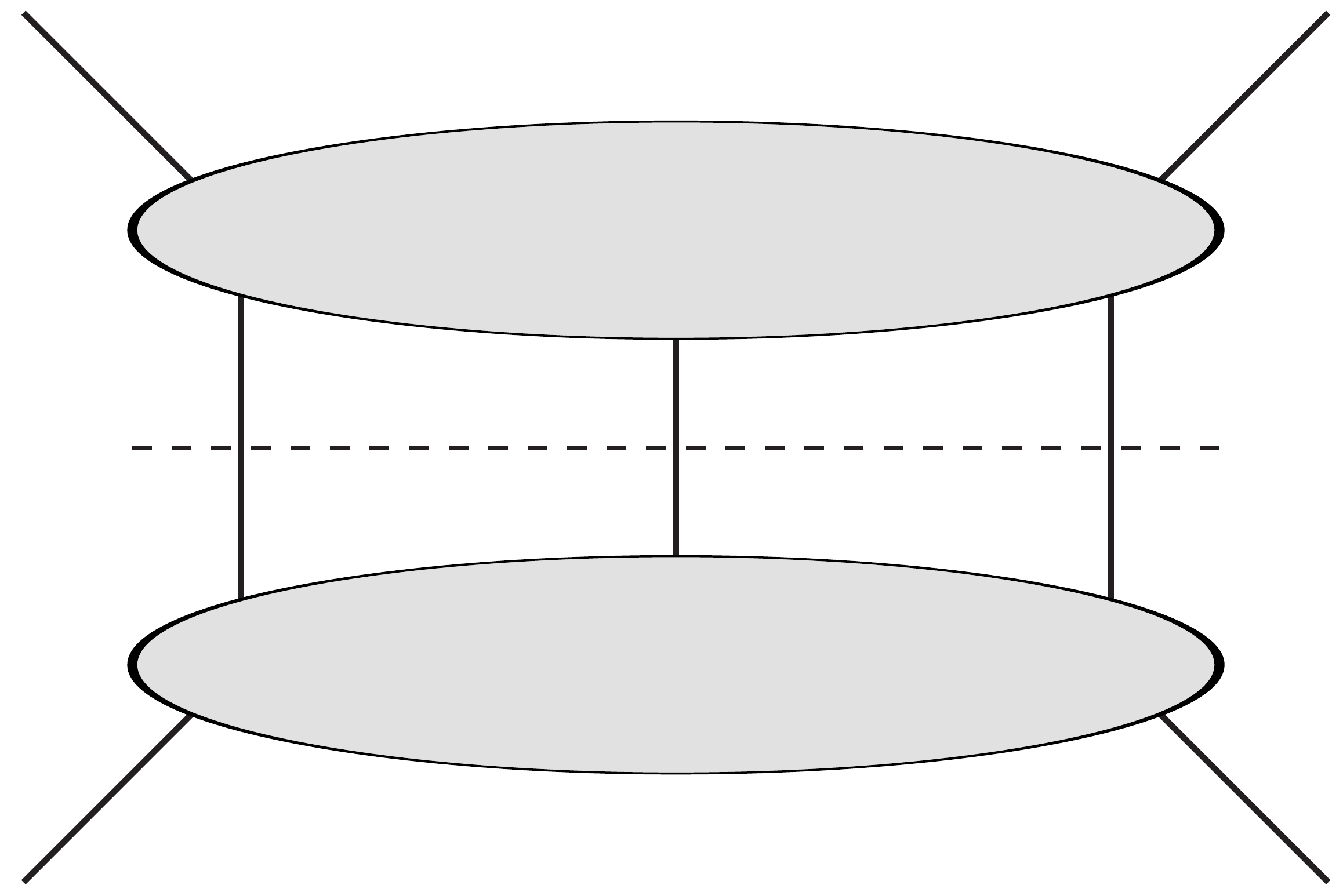}
  \caption[c]{}
\end{subfigure}
\caption{\small The three two-loop spanning cuts required at four points. The first is a butterfly cut; the 				other two are satisfied automatically by the $\mathcal{N}=4$ subsector.
         \label{fig:4ptspanningcuts}}
\end{figure}

In order to show that the BCJ presentation of the amplitude,
generated by the two masters in eq. (\ref{eq:4ptmasters}),
reproduces the full amplitude,
it is sufficient to show that it reproduces the three spanning cuts
displayed in figure \ref{fig:4ptspanningcuts}.\footnote{For
	a recent review of spanning cuts in the context of generalized unitarity, see ref.~\cite{Bern:2011qt}.}
While one could calculate these using tree-level amplitudes,
we find it simpler to start from the full presentation of the amplitude
given in ref.~\cite{Bern:2000dn}.
This can be written in terms of irreducible numerators:
\begin{align}
\Delta\bigg(\usegraph{9}{delta331l}\bigg)
&=\Delta\bigg(\!\usegraph{9}{delta322l}\bigg)
=s\,F_1(\mu_1,\mu_2),\nonumber\\
\Delta\bigg(\usegraph{9}{delta330l}\bigg)
&=F_2(\mu_1,\mu_2)+(D_s-2)^2\mu_{11}\mu_{22}\left(\frac{s+(\ell_1+\ell_2)^2}{s}\right).
\end{align}
The full colour-dressed amplitude is then
\begin{align}\label{eq:4ptamplitude}
&\mathcal{A}^{(2)}\left(1^+,2^+,3^+,4^+\right) & \nonumber\\
&=ig^6\mathcal{T}\sum_{\sigma \in S_4}\sigma\circ
I\Bigg[\frac{1}{4}C\bigg(\usegraph{9}{delta331i}\bigg)
\Bigg(\Delta\bigg(\usegraph{9}{delta331l}\bigg)
+\Delta\bigg(\usegraph{9}{delta330l}\bigg)\Bigg)\nonumber\\
&\qquad\qquad\qquad\qquad\qquad\qquad\qquad\qquad\!\!
+\frac{1}{4}C\bigg(\!\usegraph{9}{delta322i}\bigg)
\Delta\bigg(\!\usegraph{9}{delta322l}\bigg)\Bigg],
\end{align}
where for notational convenience we have redefined the integration operator $I$ to act on each $\Delta_i$ as
\begin{align}
I[\Delta_i]\equiv\int\frac{\d^d \ell_1 \d^d \ell_2}{(2\pi)^{2d}} \frac{\Delta_i}{D_i}.
\end{align}

It is known~\cite{Badger:2012dp}
that a subset of the cuts of $\mathcal{N}=4$ and all-plus two-loop amplitudes are
related by replacing the supersymmetric delta function $\delta^8(Q)$ with the function $F_1$.
Cuts involving butterfly topologies have a different structure.
This is analogous to the dimension shift between the self-dual sector and $\mathcal{N}=4$ SYM at one loop,
first observed in ref.~\cite{Bern:1996ja}.

With this in mind, we consider cuts (b) and (c).
The only diagrams contributing to these (or any) cuts are those that include all of the cut propagators.
For cuts (b) and (c),
cutting the central $(\ell_1+\ell_2)^2$ propagator ensures that
all constituent diagrams from the BCJ presentation of the amplitude belong to the $\mathcal{N}=4$ subsector,
i.e. there are no butterflies included,
so terms proportional to $(D_s-2)^2$ vanish.
Therefore,
on these cuts the all-plus colour-dual numerators are precisely equal to their $\mathcal{N}=4$ counterparts,
with the replacement $\delta^8(Q)\to F_1$.
This is consistent with the BCJ presentation of the amplitude.

Cut (a) contains butterflies, and therefore requires a little more work.
We need only consider the planar colour-stripped form:
it is known that nonplanar information is encoded in the planar cut,
as discussed by BCJ in ref.~\cite{Bern:2008qj}.\footnote{See also ref.~\cite{Badger:2015lda}.}
A simple check confirms that this cut is satisfied for terms up to a single power of $(D_s-2)$,
despite the new factors of $F_2$.
For terms proportional to $(D_s-2)^2$,
the full expression for the planar colour-stripped form of cut (a),
given in terms of irreducible numerators, is
\begin{align}\label{eq:cut220}
\left.(\ell_1-p_1)^2(\ell_2-p_5)^2\text{Cut}\bigg(\usegraph{9}{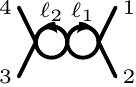}\bigg)\right|_{(D_s-2)^2}
&=\left.\Delta\bigg(\usegraph{9}{delta330l}\bigg)\right|_{(D_s-2)^2}\nonumber\\
&=\mu_{11}\mu_{22}\left(\frac{s+(\ell_1+\ell_2)^2}{s}\right),
\end{align}
where the relevant cut conditions are $\ell_1^2=\ell_2^2=(\ell_1-p_{12})^2=(\ell_2+p_{12})^2=0$.
In terms of BCJ numerators,
the only nonzero numerators that contribute are the double box, eq. (\ref{eq:n331}),
and the double triangle, eq. (\ref{eq:n3302}).
The cut in this presentation becomes
\begin{align}\label{eq:cut220bcj}
&\left.(\ell_1-p_1)^2(\ell_2-p_5)^2\text{Cut}\bigg(\usegraph{9}{delta220l}\bigg)\right|_{(D_s-2)^2}\nonumber\\
&=\left.\left(\frac{1}{(\ell_1+\ell_2)^2}n\bigg(\usegraph{9}{delta331l}\bigg)
+\frac{1}{s}n\bigg(\usegraph{9}{n330l}\bigg)\right)\right|_{(D_s-2)^2}
=\mu_{11}\mu_{22}\left(\frac{s+(\ell_1+\ell_2)^2}{s}\right),
\end{align}
as required.
At five points,
cut equations similar to this one will be important for us.

\section{The Five-Point, Two-Loop BCJ System}\label{sec:5points}

At five points and two loops,
the BCJ presentation of the $\mathcal{N}=4$  amplitude,
previously computed by Carrasco \& Johansson (CJ) in ref.~\cite{Carrasco:2011mn},
forms our starting point for the all-plus calculation.
CJ found it useful to introduce a set of prefactors $\gamma_{ij}$
that generalise the four-point prefactor $\mathcal{T}$,
given in eq. (\ref{eq:4ptprefactor}), to five points.
These prefactors encode all external state dependence on helicity:
we find them to be equally applicable to the all-plus calculation
as to the supersymmetric one.\footnote{These
	objects also prove useful for the five-point tree amplitude~\cite{Broedel:2011pd}.}

The kinematic prefactors $\gamma_{ij}$ are defined in terms of the objects $\beta_{ijklm}$ as
\begin{align}\label{eq:cjgammas}
\beta_{12345}=\frac{[12][23][34][45][51]}{4\,\epsilon(1234)},\qquad
\gamma_{12}=\beta_{12345}-\beta_{21345}=\frac{[12]^2[34][45][35]}{4\,\epsilon(1234)}
\end{align}
in the standard spinor-helicity formalism,
where $\epsilon(1234)=\epsilon_{\mu\nu\rho\sigma}p_1^\mu p_2^\nu p_3^\rho p_4^\sigma$.\footnote{In
	ref.~\cite{Carrasco:2011mn}, CJ include the supersymmetric delta function $\delta^8(Q)$ in their definition of $\beta_{12345}$; for notational convenience we include this delta function elsewhere.}
As $\gamma_{12}$ is totally symmetric on external legs $p_3$, $p_4$ and $p_5$,
we drop these last three subscripts.
The prefactors satisfy linear relations
\begin{align}
\sum_{i=1}^5\gamma_{ij}=0,\qquad
\gamma_{ij}=-\gamma_{ji},
\end{align}
so only six $\gamma_{ij}$ are linearly independent,
though they satisfy more complex relationships when kinematic factors $s_{ij}=(p_i+p_j)^2$ are involved.
We are also able to write
\begin{align}\label{eq:betagamma}
\beta_{12345}=\frac{1}{2}(\gamma_{12}+\gamma_{23}+\gamma_{13}+\gamma_{45}).
\end{align}

The all-plus pentabox numerator by itself is a master for the five-point, two-loop all-plus amplitude,
in the same way that the $\mathcal{N}=4$ pentabox is for the corresponding supersymmetric amplitude.
By analogy to the double-box numerator given in eq. (\ref{eq:n331}),
we propose that
\begin{align}\label{eq:n431}
n\bigg(\usegraph{9}{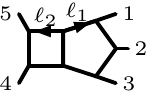}\!\bigg) = & F_1(\mu_1,\mu_2) \tilde {n}^{[\mathcal{N}=4]}\bigg(\usegraph{9}{delta431l}\!\bigg)\nonumber\\
&\qquad
+(D_s-2)^2\mu_{11}\mu_{22}(\ell_1+\ell_2)^2X(12345;\ell_1,\ell_2),
\end{align}
where $X$ is an unknown function of external and loop momenta  which we must determine,
while $\tilde {n}^{[\mathcal{N}=4]}$ is the coefficient of the supersymmetric delta function in CJ's $\mathcal{N}=4$ supersymmetric pentabox numerator:
\begin{align}\label{eq:n431N4}
n^{[\mathcal{N}=4]}\bigg(\usegraph{9}{delta431l}\!\bigg)& = \delta^{(8)}(Q) \tilde {n}^{[\mathcal{N}=4]}\bigg(\usegraph{9}{delta431l}\!\bigg) \nonumber\\
=\frac{1}{4}\delta^8(Q)\big(&\gamma_{12}(2s_{45}-s_{12}+2\ell_1\cdot(p_2-p_1))
+\gamma_{23}(s_{45}+2s_{12}+2\ell_1\cdot(p_3-p_2))\nonumber\\
+4&\gamma_{45}\,\ell_1\cdot(p_5-p_4)
+\gamma_{13}(s_{12}+s_{45}+2\ell_1\cdot(p_3-p_1))\big).
\end{align}
For notational convenience below, we will drop the tilde on $\tilde {n}^{[\mathcal{N}=4]}$.

The all-plus pentabox numerator shares many of the properties of the all-plus double box.
There is a non-closed $\mathcal{N}=4$ subsector generated by terms containing up to one power of $(D_s-2)$,
where once again the extra-dimensional function $F_1$, given in eq. (\ref{eq:F1}),
plays the role of the supersymmetric delta function $\delta^8(Q)$.
Jacobi relations are again divided into two categories:
those that preserve both $F_1$ and $(\ell_1+\ell_2)^2$,
and those that act upon them.
The unknown function $X$ gives us information about butterfly topologies.
However, unlike the situation at four points,
we will show that $X$ is necessarily nonlocal in external kinematics.

\subsection{Symmetries and automorphisms}
\label{sec:5ptsymmetry}

We choose to recycle another desirable property of the four-point solution:
the absence of terms proportional to $(D_s-2)^2$ in all nonplanar numerators.
Although there is no five-point equivalent of the four-point nonplanar master, eq. (\ref{eq:n322}),
we can still impose this condition using the following nonplanar numerator:
\begin{align}\label{eq:n332}
n\bigg(&\usegraph{9}{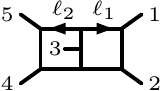}\bigg)
=n\bigg(\usegraph{9}{delta431l}\!\bigg)-n\bigg(\!\usegraph{9}{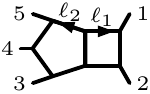}\bigg)
=F_1(\mu_1,\mu_2)n^{[\mathcal{N}=4]}\bigg(\usegraph{9}{delta332l}\bigg) \nonumber \\
&+(D_s-2)^2\mu_{11}\mu_{22}(\ell_1+\ell_2)^2
(X(12345;\ell_1,\ell_2)-X(34512;-p_{12}-\ell_2,p_{12}-\ell_1)).
\end{align}

All nonplanar numerators at five points can be generated from this graph alone using Jacobi identities.
In this sense, it is a master of nonplanar graphs.
Therefore, vanishing of terms proportional to $(D_s-2)^2$ in this numerator is necessary and sufficient
to guarantee the same property for all nonplanar graphs.
Imposing this requirement, we learn that
$X(12345;\ell_1,\ell_2)=X(34512;-p_{12}-\ell_2,p_{12}-\ell_1)$.

Another property of $X$ follows from the overall flip symmetry of the pentabox through a horizontal axis,
namely $X(12345;\ell_1,\ell_2)=-X(32154;-p_{45}-\ell_1,p_{45}-\ell_2)$.
Both of these properties are functional identities,
holding for any permutation of external legs and any shift in the loop momenta.
By applying the two conditions to each other repeatedly and performing relabellings,
we can refine them into three simple properties of $X$:
\begin{align}\label{eq:Xsymmetries}
X(12345;\ell_1,\ell_2)=\left\{ 
  \begin{array}{l}
    X(23451;\ell_1-p_1,\ell_2+p_1), \\
    -X(54321;\ell_2,\ell_1), \\ 
    X(12345;-\ell_2,-\ell_1).
  \end{array} \right. 
\end{align}
These identities will be important below.\footnote{At four points, the
	double box satisfies an analogous set of symmetries, with the constant $1/2$ playing the role of $X$. As the number of vertices (six) is even, the second symmetry is in this case positive.}

We have exhaustively checked that these three conditions are sufficient to guarantee
all symmetries and automorphisms for all descendent BCJ numerators in the entire system.
For instance, the nonplanar numerator in eq. (\ref{eq:n332}) now equals its $\mathcal{N}=4$ term alone,
so it satisfies its symmetries by virtue of the $\mathcal{N}=4$ numerator having exactly the same properties.

From the planar sector,
a more nontrivial example of these symmetries in action comes from the ``hexatriangle'' diagram:
\begin{align}\label{eq:n521}
n\bigg(\!\!\usegraph{13}{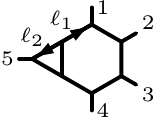}\bigg)
&=n\bigg(\usegraph{9}{delta431l}\!\bigg)-n\bigg(\!\usegraph{9}{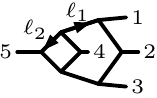}\!\bigg)\nonumber\\
&=(D_s-2)^2\mu_{11}\mu_{22}(\ell_1+\ell_2)^2X(12345;\ell_1,\ell_2).
\end{align}
Here we have used the fact that
\begin{align}\label{eq:N4planarnonplanar}
n^{[\mathcal{N}=4]}\bigg(\usegraph{9}{delta431l}\!\bigg)
=n^{[\mathcal{N}=4]}\bigg(\!\usegraph{9}{delta422l}\!\bigg),
\end{align}
and that the nonplanar numerator in eq. (\ref{eq:n521}) has no terms carrying powers of $(D_s-2)^2$.
As expected, the hexatriangle has no $\mathcal{N}=4$ component (it contains an internal triangle).
Its symmetry through a horizontal axis implies
$X(12345;\ell_1,\ell_2)=-X(43215;-p_5-\ell_1,p_5-\ell_2)$,
which follows from the three conditions in eq. (\ref{eq:Xsymmetries}).
The Jacobi identity in eq. (\ref{eq:n521}) partitions the pentabox into its $\mathcal{N}=4$
and pure YM components.

\subsection{Spanning cuts}
\label{sec:5ptcuts}

\begin{figure}[t]
\centering
\begin{subfigure}{0.25\textwidth}
  \centering
  \includegraphics[width=0.8\textwidth]{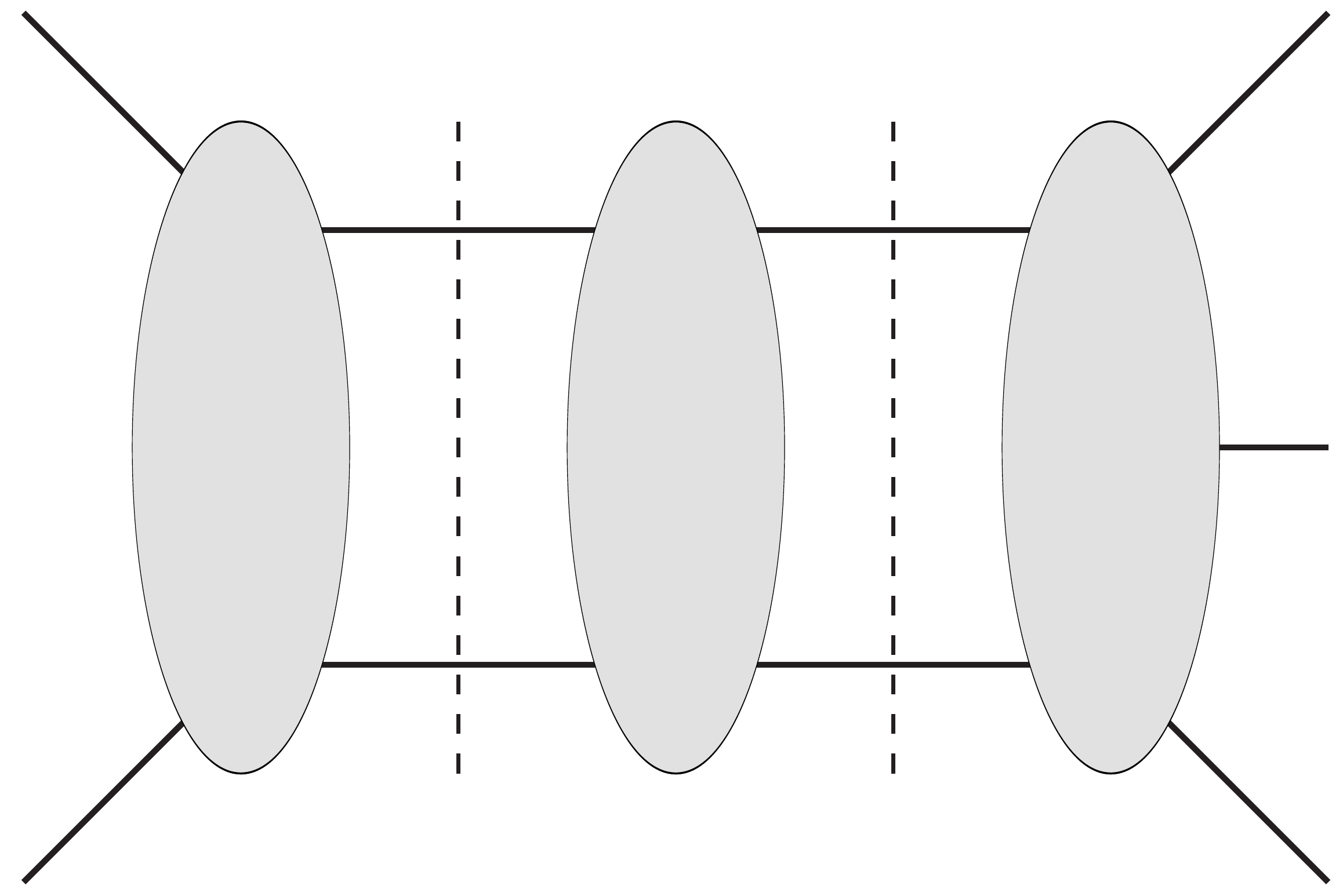}
  \caption[a]{}
\end{subfigure}
\begin{subfigure}{0.25\textwidth}
  \centering
  \includegraphics[width=0.8\textwidth]{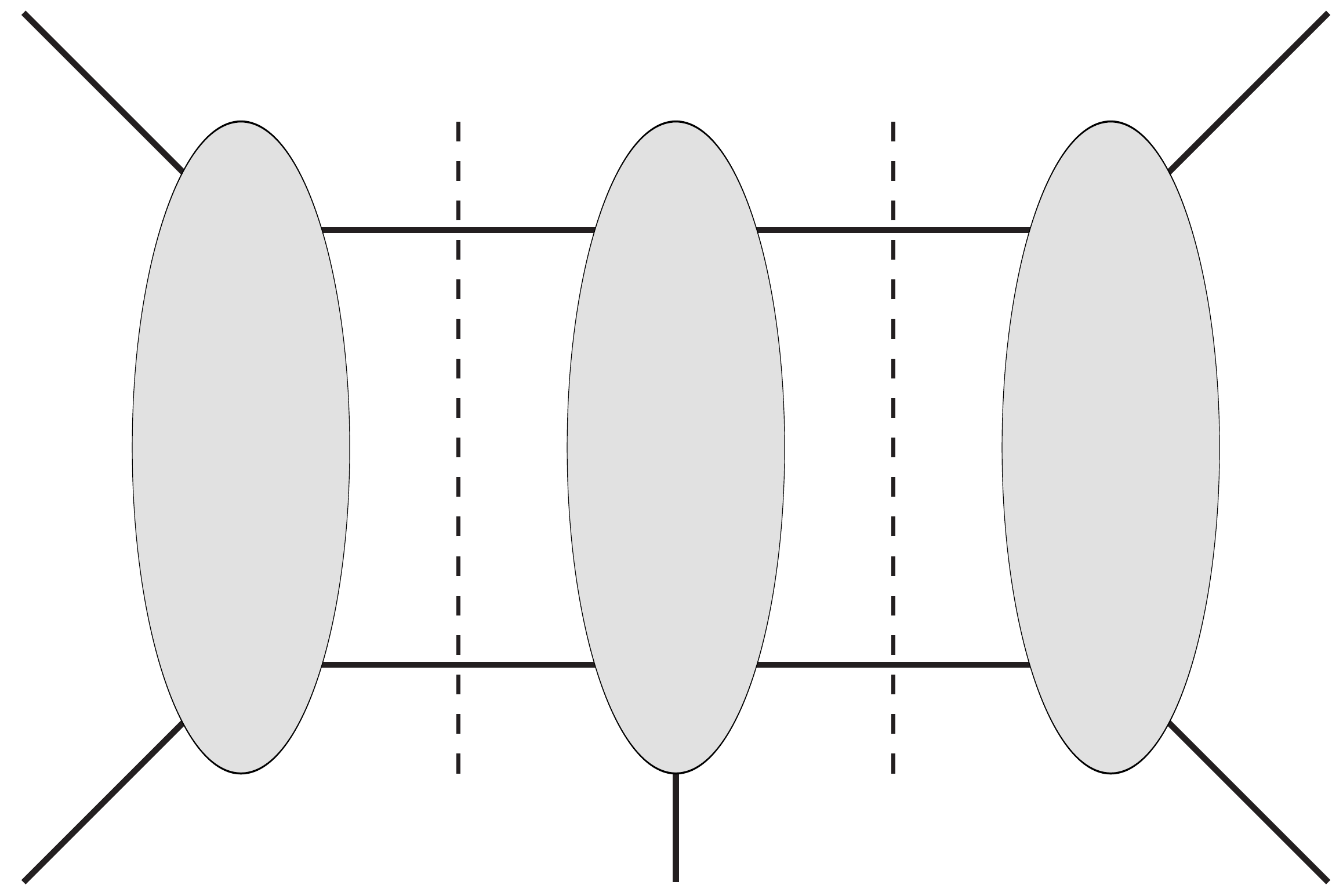}
  \caption[b]{}
\end{subfigure}
\begin{subfigure}{0.25\textwidth}
  \centering
  \includegraphics[width=0.8\textwidth]{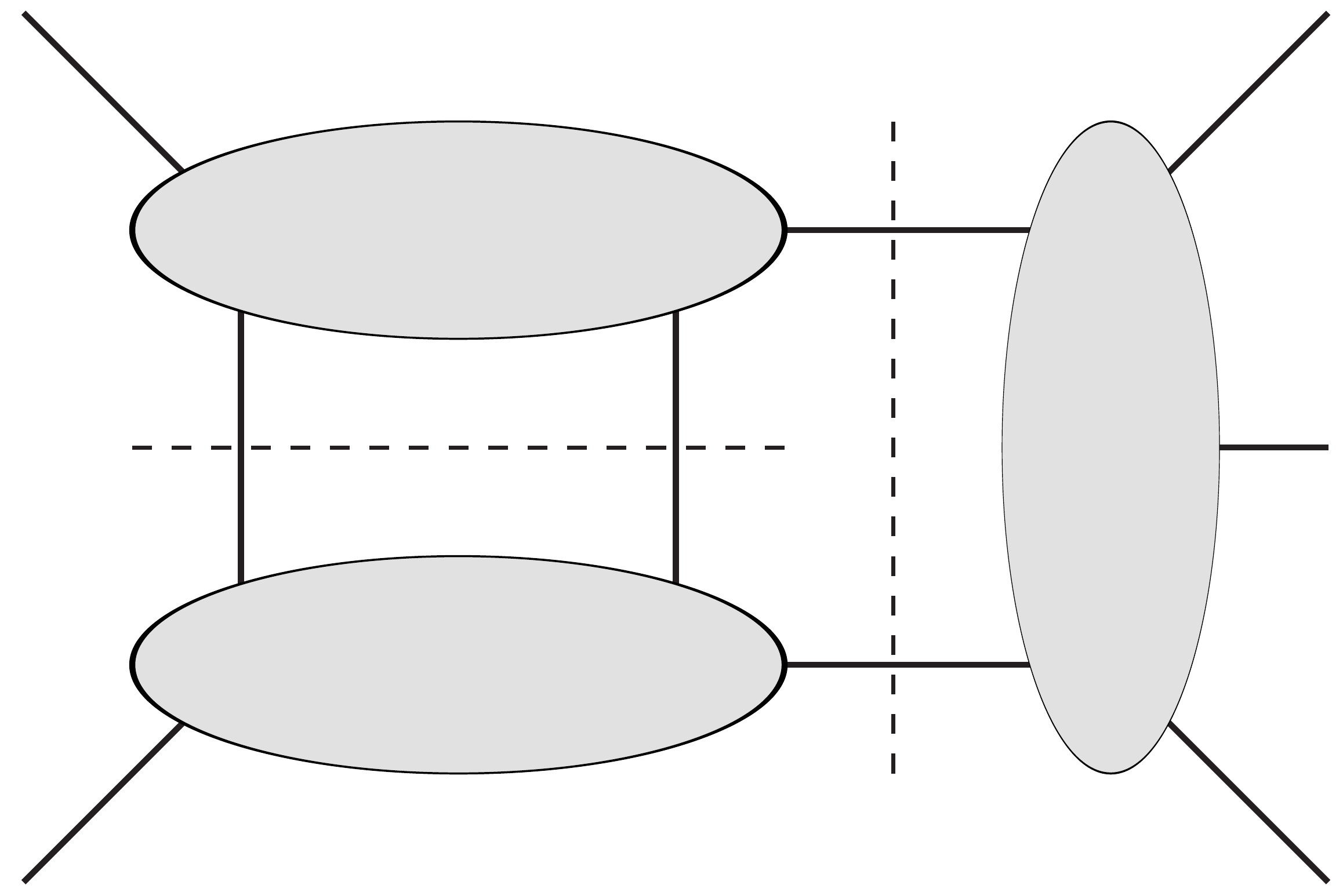}
  \caption[c]{} 
\end{subfigure}
\begin{subfigure}{0.25\textwidth}
  \centering
  \includegraphics[width=0.8\textwidth]{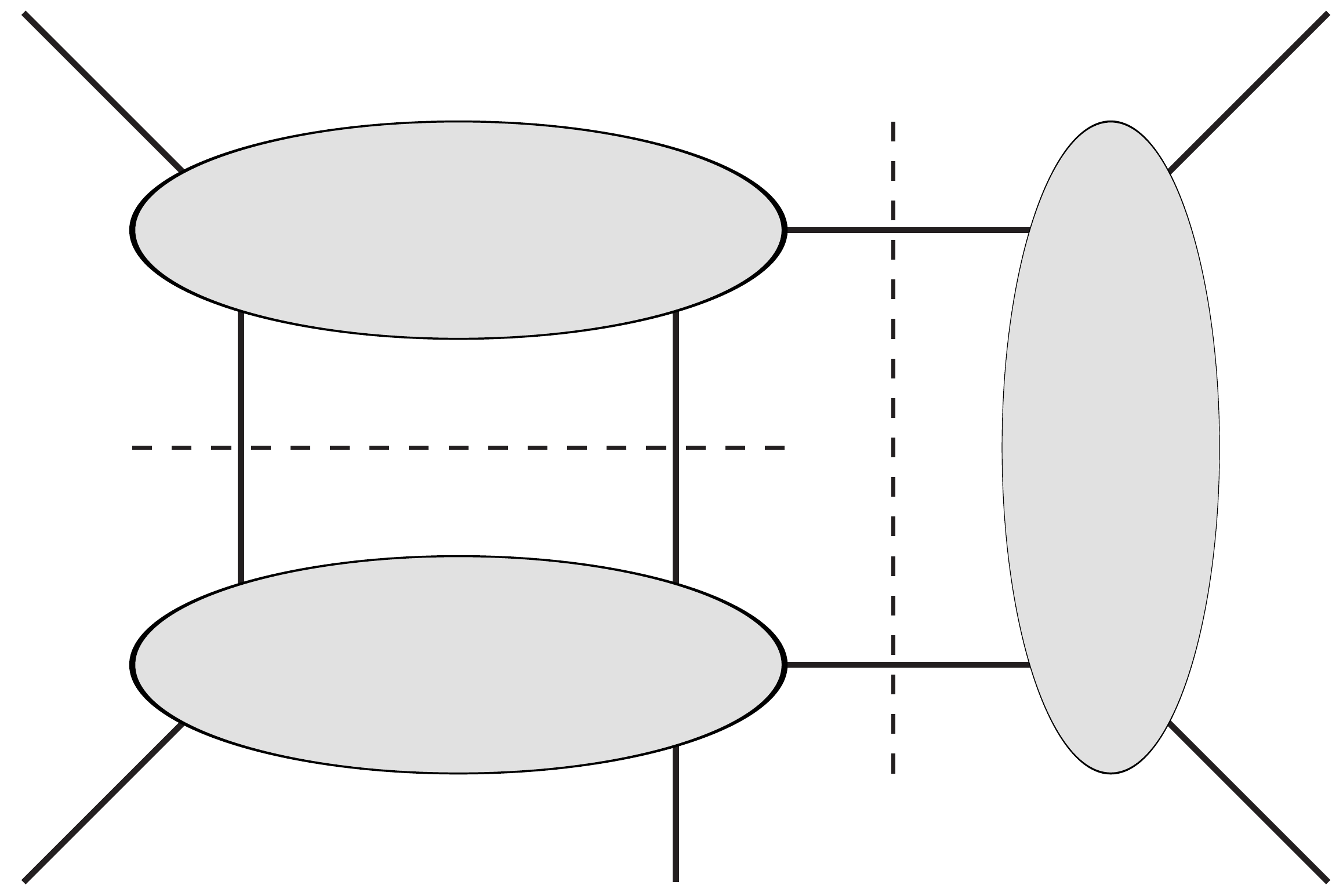}
  \caption[d]{}
\end{subfigure}
\begin{subfigure}{0.25\textwidth}
  \centering
  \includegraphics[width=0.8\textwidth]{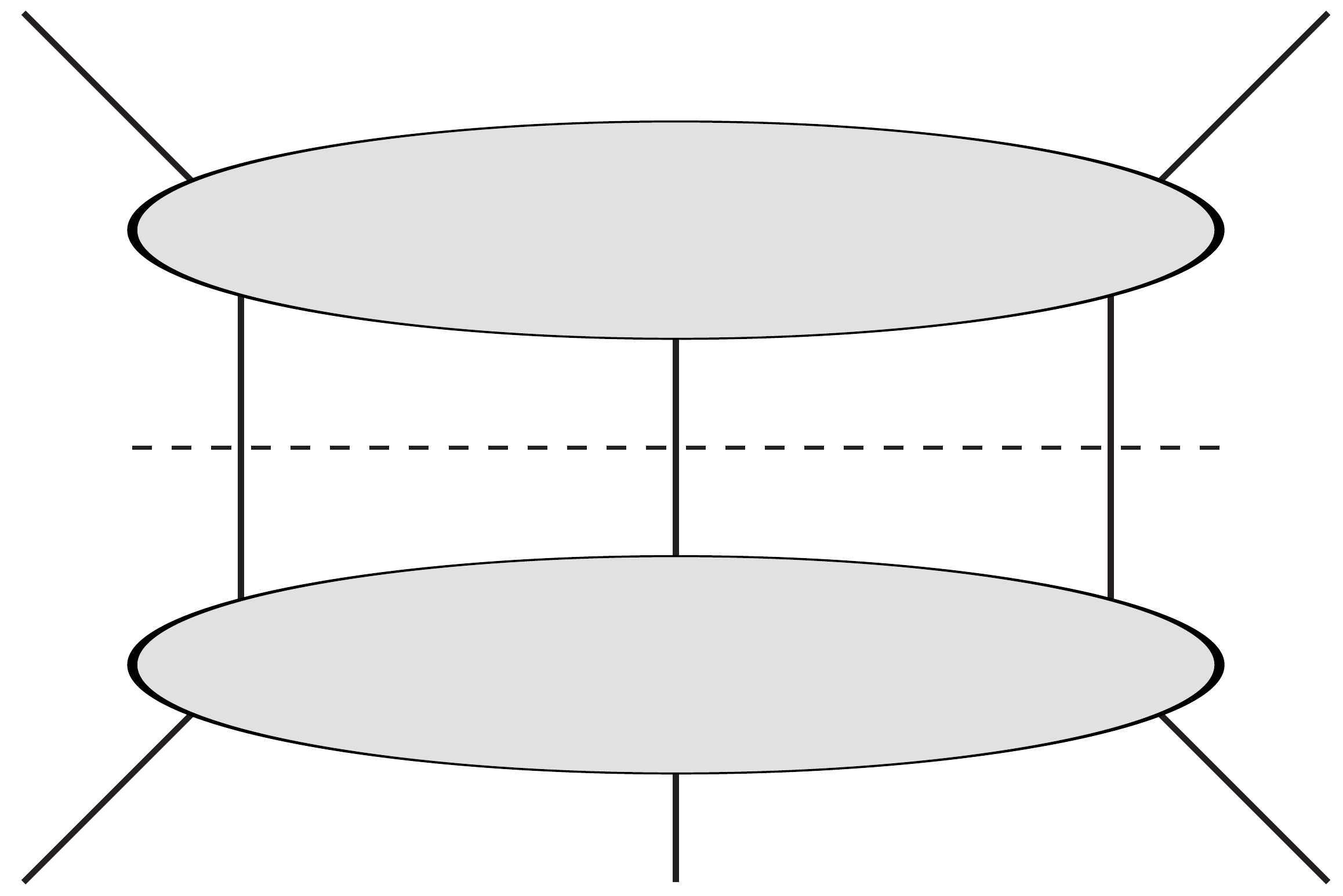}
  \caption[e]{}
\end{subfigure} 
\begin{subfigure}{0.25\textwidth}
  \centering
  \includegraphics[width=0.8\textwidth]{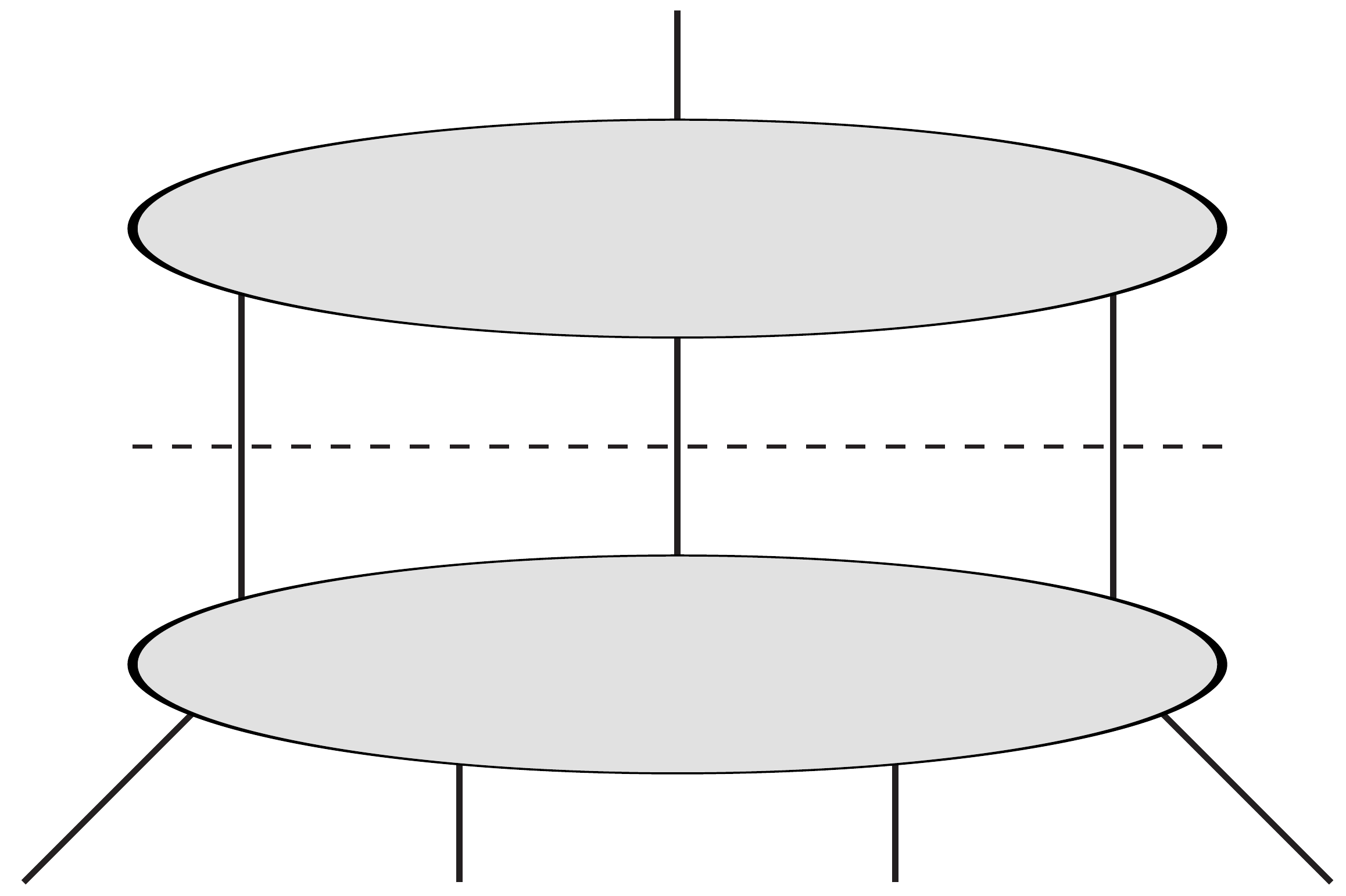}
  \caption[f]{}
\end{subfigure}
\caption{\small A spanning set of cuts at  two loops and five points. The first two are butterfly cuts; the rest are satisfied automatically by the $\mathcal{N}=4$ subsector.
         \label{fig:5ptspanningcuts}}
\end{figure}

To determine the unknown function $X$, 
we now match to physical information using a set of spanning cuts,
displayed in figure \ref{fig:5ptspanningcuts}.
We compare to the irreducible presentation of the amplitude computed by Badger,
Frellesvig and Zhang (BFZ) ~\cite{Badger:2013gxa}.
As in the four points example, 
spanning cuts that involve cutting the central $(\ell_1+\ell_2)^2$ propagator
do not provide any information on $X$:
in this case, the first two cuts in figure \ref{fig:5ptspanningcuts} contain the relevant new physics.
Again, it suffices to reproduce them in the planar sector as the nonplanar sector follows through
tree-level relations on colour-ordered amplitudes.

The first of these two cuts, using the BFZ amplitude, is
\begin{align}\label{eq:cut2205BFZ}
&(\ell_1-p_1)^2(\ell_1-p_{12})^2(\ell_2-p_5)^2
\left.\text{Cut}\bigg(\usegraph{9}{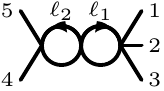}\bigg)\right|_{(D_s-2)^2\mu_{11}\mu_{22}}\nonumber\\
&=\frac{s_{45}+(\ell_1+\ell_2)^2}{s_{45}}\left(\beta_{12345}
+\frac{\gamma_{45}}{s_{45}}(\ell_1+p_5)^2
+\frac{\gamma_{12}}{s_{12}}(\ell_1-p_1)^2
+\frac{\gamma_{23}}{s_{23}}(\ell_1-p_{12})^2\right).
\end{align}
We may also evaluate this cut using the BCJ master numerator, eq. \eqref{eq:n431}. For brevity,
we define $X'(12345;\ell_1,\ell_2)\equiv X(12345;\ell_1,\ell_2)+X(12354;\ell_1,p_{45}-\ell_2)$
and we also introduce antisymmetric brackets,
e.g. $X'([12]345;\ell_1,\ell_2)=X'(12345;\ell_1,\ell_2)-X'(21345;\ell_1,\ell_2)$. Using this
notation, and the on-shell conditions $\ell_1^2=(\ell_1+p_{45})^2=\ell_2^2=(\ell_2-p_{45})^2=0$, we find
\begin{align}\label{eq:cut2205BCJ}
&(\ell_1-p_1)^2(\ell_1-p_{12})^2(\ell_2-p_5)^2
\left.\text{Cut}\bigg(\usegraph{9}{delta2205l}\bigg)\right|_{(D_s-2)^2\mu_{11}\mu_{22}}\nonumber\\
&=\frac{s_{45}+(\ell_1+\ell_2)^2}{s_{45}}\bigg(X'(12345;\ell_1,\ell_2)
+\frac{X'([12]345;\ell_1,\ell_2)}{s_{12}}(\ell_1-p_1)^2
\nonumber\\
&+\frac{X'(1[23]45;\ell_1,\ell_2)}{s_{23}}(\ell_1-p_{12})^2
+\frac{X'(123[45];\ell_1,\ell_2)}{s_{45}}(\ell_2-p_5)^2
\nonumber\\
&+\frac{X'([[12]3]45;\ell_1,\ell_2)}{s_{12}s_{45}}(\ell_1-p_1)^2(\ell_1-p_{12})^2
+\frac{X'([1[23]]45;\ell_1,\ell_2)}{s_{23}s_{45}}(\ell_1-p_1)^2(\ell_1-p_{12})^2
\nonumber\\
&+\frac{X'([12]3[45];\ell_1,\ell_2)}{s_{12}s_{45}}(\ell_1-p_1)^2(\ell_2-p_5)^2
+\frac{X'(1[23][45];\ell_1,\ell_2)}{s_{23}s_{45}}(\ell_1-p_{12})^2(\ell_2-p_5)^2
\nonumber\\
&+\frac{X'([[12]3][45];\ell_1,\ell_2)}{s_{12}s_{45}^2}(\ell_1-p_1)^2(\ell_1-p_{12})^2(\ell_2-p_5)^2
\nonumber\\
&+\frac{X'([1[23]][45];\ell_1,\ell_2)}{s_{23}s_{45}^2}(\ell_1-p_1)^2(\ell_1-p_{12})^2(\ell_2-p_5)^2\bigg).
\end{align}

This cut bears a strong similarity to its four-point counterpart,
given in eqs. (\ref{eq:cut220}) and (\ref{eq:cut220bcj}),
preserving an overall flip symmetry through the horizontal axis.
The prefactor $\mathcal{T}$ has now been generalised to the five-point kinematic prefactors $\gamma_{ij}$.

The other planar cut has a somewhat more complicated structure. Using the BFZ presentation of the amplitude,
we find
\begin{align}\label{eq:cut2205LBFZ}
&(\ell_1-p_1)^2(\ell_2-p_5)^2(\ell_1+p_{45})^2(\ell_2+p_{12})^2
\left.\text{Cut}\bigg(\usegraph{9}{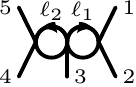}\bigg)\right|_{(D_s-2)^2\mu_{11}\mu_{22}}
\nonumber\\
&=(\ell_2+p_{12})^2\left(\frac{s_{45}+(\ell_1+\ell_2)^2}{s_{45}}\right)
\left(\beta_{12345}+\frac{\gamma_{45}}{s_{45}}(\ell_1+p_5)^2+\frac{\gamma_{12}}{s_{12}}(\ell_1-p_1)^2\right)
\nonumber\\
&\;\;\;\;\;\;
+\frac{(\ell_1+p_{45})^2(\ell_2+p_{12})^2}{s_{12}s_{45}}\bigg[
\frac{\beta_{12345}}{2}(\ell_1+\ell_2)^2\nonumber\\
&\;\;\;\;\;\;\;\;\;\;\;\;
+\left(s_{45}-s_{12}+(\ell_2+p_{12})^2\right)
\left(\frac{\gamma_{45}}{s_{45}}(\ell_1+p_5)^2+\frac{\gamma_{12}}{s_{12}}(\ell_1-p_1)^2\right)
\nonumber\\
&\;\;\;\;\;\;\;\;\;\;\;\;
-\left(s_{12}(\gamma_{12}+\gamma_{45}-\beta_{12345})+s_{13}\gamma_{12}\right)\times
\nonumber\\
&\;\;\;\;\;\;\;\;\;\;\;\;\;\;\;\;\;\;
\left(\frac{(\ell_1+\ell_2)^2}{s_{12}}
+\frac{1}{2}\left(\frac{s_{45}-(\ell_1+p_{45})^2}{s_{45}}\right)
\left(\frac{s_{12}-(\ell_2+p_{12})^2}{s_{12}}\right)\right)\bigg]
\nonumber\\
&\;\;\;\;\;\;
-(p_1\leftrightarrow p_5,p_2\leftrightarrow p_4,\ell_1\leftrightarrow\ell_2).
\end{align}
Meanwhile, using the BCJ presentation of the amplitude, we find
\begin{align}\label{eq:cut2205LBCJ}
&(\ell_1-p_1)^2(\ell_2-p_5)^2(\ell_1+p_{45})^2(\ell_2+p_{12})^2
\left.\text{Cut}\bigg(\usegraph{9}{delta2205Ll}\bigg)\right|_{(D_s-2)^2\mu_{11}\mu_{22}}
\nonumber\\
&=\bigg\{(\ell_2+p_{12})^2\left(\frac{s_{45}+(\ell_1+\ell_2)^2}{s_{45}}\right)
\bigg(X'(12345;\ell_1,\ell_2)+\frac{X'([12]345;\ell_1,\ell_2)}{s_{12}}(\ell_1-p_1)^2
\nonumber\\
&\;\;\;\;\;\;
+\frac{X'(123[45];\ell_1,\ell_2)}{s_{45}}(\ell_2-p_5)^2
+\frac{X'([12]3[45];\ell_1,\ell_2)}{s_{12}s_{45}}(\ell_1-p_1)^2(\ell_2-p_5)^2\bigg)
\nonumber\\
&\;\;\;\;\;\;\;\;\;\;\;\;
-(p_1\leftrightarrow p_5,p_2\leftrightarrow p_4,\ell_1\leftrightarrow\ell_2)\bigg\}\nonumber\\
&\;\;\;\;\;\;
+\frac{(\ell_1+p_{45})^2(\ell_2+p_{12})^2}{s_{12}s_{45}}\bigg(
X''(12345;\ell_1,\ell_2)
+\frac{X''([12]345;\ell_1,\ell_2)}{s_{12}}(\ell_1-p_1)^2
\nonumber\\
&\;\;\;\;\;\;\;\;\;\;\;\;
+\frac{X''(123[45];\ell_1,\ell_2)}{s_{45}}(\ell_2-p_5)^2
+\frac{X''([12]3[45];\ell_1,\ell_2)}{s_{12}s_{45}}(\ell_1-p_1)^2(\ell_2-p_5)^2\bigg),
\end{align}
where this time we have used the on-shell conditions
$\ell_1^2=(\ell_1-p_{12})^2=\ell_2^2=(\ell_2-p_{45})^2=0$,
and also the three symmetries on $X$ introduced in eq. (\ref{eq:Xsymmetries}).
We have also introduced
\begin{align}\label{eq:xdoubleprimed}
&X''(12345;\ell_1,\ell_2)\equiv(\ell_1+\ell_2)^2X(12345;\ell_1,\ell_2)\nonumber\\
&\qquad
+(s_{45}-s_{12}+(\ell_1+\ell_2)^2-(\ell_1+p_{45})^2)X(12354;\ell_1,p_{45}-\ell_2)\nonumber\\
&\qquad
+(s_{12}-s_{45}+(\ell_1+\ell_2)^2-(\ell_2+p_{12})^2)X(21345;p_{12}-\ell_1,\ell_2)\nonumber\\
&\qquad
+((\ell_1+\ell_2)^2-(\ell_1+p_{45})^2-(\ell_2+p_{12})^2)X(21354;p_{12}-\ell_1,p_{45}-\ell_2).
\end{align}
This cut preserves a symmetry through the vertical axis.

Our task is now to find a solution for $X$ that satisfies these two cut equations,
while at the same time enjoying the off-shell symmetries in eq. (\ref{eq:Xsymmetries}).

\subsection{Nonlocality properties}
\label{sec:nonlocal}

Before we describe our strategy for finding $X$, let us pause to comment on what kind
of structure we may expect. We claim that $X$ must be 
nonlocal in the kinematic factors $s_{ij}$. 
The basis of our claim is simple.
We input physical information into our calculation using the irreducible set of numerators
computed in ref.~\cite{Badger:2013gxa}, and
we further insist on writing our results in terms of the prefactors $\gamma_{ij}$.
Consider making a local ansatz for the full pentabox numerator
in terms of the six linearly independent kinematic prefactors $\gamma_{ij}$:
\begin{align}\label{eq:localansatz}
n\bigg(\usegraph{9}{delta431l}\!\bigg)
=\gamma_{12}m_1+\gamma_{13}m_2+\gamma_{14}m_3+\gamma_{23}m_4+\gamma_{24}m_5+\gamma_{34}m_6,
\end{align}
where dimensional analysis tells us that the local objects $m_i$ may carry up to six powers of loop momentum,
including factors of extra-dimensional components $\mu_{ij}$.
As the pentabox is the master for the five-point, two-loop system,
linearity of the Jacobi identities implies that \emph{all} numerators
are limited to six powers of loop momentum.

BFZ's box-triangle irreducible numerator is given by
\begin{align}\label{eq:delta430}
\Delta\bigg(\usegraph{13}{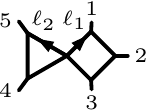}\!\!\!\:\bigg)
&=-\frac{s_{12}\text{tr}_+(1345)}{2 \SpDenom5 s_{13}}(2(\ell_1\!\cdot\!\omega_{123})+s_{23})\nonumber\\
& \qquad \qquad \,\, \times\left( F_2(\mu_1,\mu_2)+(D_s-2)^2\mu_{11}\mu_{22}
\frac{s_{45}+(\ell_1+\ell_2)^2}{s_{45}} \right);
\end{align}
it has a maximum of seven powers of loop momentum.
This continues to be true even if we move terms to other irreducible graphs with
fewer propagators using integrand reduction.
It is also still true if we choose to rewrite the regulator components $\mu_{ij}$ in terms of
conventional scalar products $\ell_i\cdot p_j$ and $\ell_i\cdot\ell_j$.
We conclude that $X$ must be a nonlocal object. Notice that our argument applies to master
numerators with a more general structure than we assumed in eq.~\eqref{eq:n431}; we merely used the
BFZ irreducible numerators in combination with an ansatz which is linear in $\gamma_{ij}$.

\subsection{Attempting a minimal solution}

Our first attempt at a solution for $X$ is motivated by power counting from the Feynman rules.
As all diagrams have seven vertices,
one would expect up to seven powers of loop momentum in a solution for the pentabox numerator:
the factor of $\mu_{11}\mu_{22}(\ell_1+\ell_2)^2$ in eq. (\ref{eq:n431}) already accounts for six,
so we anticipate $X$ having a single power of loop momentum.
Although this proposal will not prove successful,
the resulting calculation conveniently demonstrates the obstacle to colour-kinematics duality.
We shall also use it to introduce concepts and notation that will help us find a solution for $X$
in the next section.

Dependence on the two $d$-dimensional loop momenta, $\ell_i=(\bar{\ell}_i,\mu_i)$,
accounts for a total of eleven degrees of freedom:
eight coming from the components of $\bar{\ell}_1$ and $\bar{\ell}_2$,
plus an extra three from $\mu_{11}$, $\mu_{22}$ and $\mu_{12}$.
This freedom can be parametrised by eleven independent scalar products $z_i$:
\begin{align}\label{eq:z}
z_1&=\ell_1^2,  & z_2&=(\ell_1-p_1)^2,  & z_3&=(\ell_1-p_{12})^2, & z_4&=(\ell_1+p_{45})^2,\nonumber\\
z_5&=\ell_2^2,  & z_6&=(\ell_2-p_5)^2,  & z_7&=(\ell_2-p_{45})^2, & z_8&=(\ell_1+\ell_2)^2,\nonumber\\
z_9&=(\ell_1+p_5)^2,  & z_{10}&=(\ell_2+p_1)^2, & z_{11}&=(\ell_2+p_{12})^2,
\end{align}
of which the first eight are the propagators of the pentabox.
Any polynomial expression depending on scalar products of the form $\ell_i\cdot\ell_j$,
$\ell_i\cdot p_j$ or $\mu_{ij}$ can be uniquely expressed in terms of these objects,
e.g. $\ell_1\cdot\ell_2=(z_8-z_1-z_5)/2$.
This particular choice is convenient because the $z_i$ always transform directly
into each other under the symmetries in eq. (\ref{eq:Xsymmetries}).
For instance,
under the third of these symmetries which acts as $\ell_1\leftrightarrow-\ell_2$,
\begin{align}\label{eq:zmap}
z_1\leftrightarrow z_5, \qquad z_2\leftrightarrow z_{10}, \qquad z_3\leftrightarrow z_{11},
\qquad z_4\leftrightarrow z_7, \qquad z_6\leftrightarrow z_9,
\end{align}
and $z_8$ remains unchanged.

To find a solution for $X$ with minimal dependence on loop momentum,
we make an ansatz of the form
\begin{align}\label{eq:minansatz}
X(12345;\ell_1,\ell_2)=A(12345)+\sum_{i=1}^{11}A_i(12345)z_i,
\end{align}
where the unknown new functions $A$ and $A_i$ are functions of the external momenta $p_i$.
They are ``relabelling friendly'',
in the sense that knowledge of a function acting on a given ordering of its arguments
suffices to reproduce any other ordering through relabelling of the external momenta $p_i$.
However, for the reasons outlined in section \ref{sec:nonlocal},
we do not expect these functions to be local functions of external momenta.

With this setup, solving the three symmetries in eq. (\ref{eq:Xsymmetries}) is straightforward.
If one substitutes the ansatz presented in eq. (\ref{eq:minansatz}),
matching individual powers of $z_i$ gives a set of simple functional identities,
e.g. $A_1(23451)=A_2(12345)=A_{10}(12345)$, etc.
The relabelling property ensures the validity of these identities for any ordering of the external momenta,
so such identities can be used to eliminate unnecessary functions $A_i$ entirely.
Once all such identities are solved,
\begin{align}\label{eq:minsoln}
&X(12345;\ell_1,\ell_2)=A(12345)+A_1(12345)(\ell_1^2+\ell_2^2)\nonumber\\
&\qquad
+A_1(23451)((\ell_1-p_1)^2+(\ell_2+p_1)^2)+A_1(34512)((\ell_1-p_{12})^2+(\ell_2+p_{12})^2)\nonumber\\
&\qquad
+A_1(45123)((\ell_1+p_{45})^2+(\ell_2-p_{45})^2)+A_1(51234)((\ell_1+p_5)^2+(\ell_2-p_5)^2)\nonumber\\
&\qquad
+A_8(12345)(\ell_1+\ell_2)^2,
\end{align}
along with the additional requirements that
\begin{subequations}\label{eq:minsymmetries}
\begin{align}
A(12345)&=-A(54321)=A(23451),\label{eq:Asymmetry}\\
A_1(12345)&=-A_1(54321),\label{eq:A1symmetry}\\
A_8(12345)&=-A_8(54321)=A_8(23451).\label{eq:A8symmetry}
\end{align}
\end{subequations}

It now suffices to consider a box-triangle cut,
an expression for which may be deduced from either of the
two double-bubble cuts in section \ref{sec:5ptcuts}.
The cut conditions are
$\ell_1^2=(\ell_1-p_1)^2=(\ell_1-p_{12})^2=(\ell_1+p_{45})^2=\ell_2^2=(\ell_2-p_5)^2=(\ell_2-p_{45})^2=0$,
yielding a condition on $X$:
\begin{align}\label{eq:cut430}
\left.\text{Cut}\bigg(\usegraph{13}{delta430l}\!\!\!\:\bigg)\right|_{(D_s-2)^2\mu_{11}\mu_{22}}
&=\frac{s_{45}+(\ell_1+\ell_2)^2}{s_{45}}\left(\beta_{12345}
+\frac{\gamma_{45}}{s_{45}}(\ell_1+p_5)^2\right)\nonumber\\
&=\frac{s_{45}+(\ell_1+\ell_2)^2}{s_{45}}X'(12345;\ell_1,\ell_2),
\end{align}
where $X'(12345;\ell_1,\ell_2)\equiv X(12345;\ell_1,\ell_2)+X(12354;\ell_1,p_{45}-\ell_2)$ as before.
Rewriting this in terms of the new variables $z_i$,
on which the cut conditions are $z_i=0$ for $i\leqslant7$,
\begin{align}
\alpha_{12345}+\frac{\gamma_{45}}{s_{45}}z_9
&=A(123\{45\})+s_{12}A_1(35412)+s_{23}A_1(23541)\nonumber\\
&\qquad-s_{45}(A_1(35412)+A_1(41235)+A_1(23541)+A_8(12354))\nonumber\\
&\qquad+A_8(123[45])z_8+(A_1(51234)-A_1(41235))z_9\nonumber\\
&\qquad+A_1(23[45]1)z_{10}+A_1(3[45]12)z_{11}.
\end{align}
Here we have chosen to explore a more general equation
with the unknown object $\alpha_{12345}$ playing the role of $\beta_{12345}$.
Symmetry of the box-triangle cut implies that $\alpha_{12345}=-\alpha_{32154}$.
We have also introduced an anticommutator, $A(123\{45\})\equiv A(12345)+A(12354)$.

We proceed by matching powers of the four nonzero scalar products $z_i$.
The coefficient of $z_8$ reveals that $A_8(123[45])=0$;
together with the requirement from eq. (\ref{eq:A8symmetry}) that $A_8(12345)=A_8(23451)$,
we see that $A_8$ is totally permutation symmetric on its five arguments.
The antisymmetry property in eq. (\ref{eq:A8symmetry}) therefore requires $A_8$ to vanish.\footnote{This
	is to be expected: were $A_8$ nonzero,	it would necessitate more than one power of loop momentum in the final solution for $X$ as $\ell_1\cdot\ell_2$ cannot be cancelled elsewhere.}
The coefficients of $z_{10}$ and $z_{11}$ tell us that $A_1(1[23]45)=A_1(12[34]5)=0$.
The coefficient of $z_9$ then provides a unique solution:
\begin{align}\label{eq:A1}
A_1(12345)=\frac{\gamma_{51}}{2s_{15}}, \qquad A_8(12345)=0.
\end{align}

The condition one would use to solve for $A$, with zero powers of $z_i$, presents an obstacle.
Substituting the results for $A_1$ and $A_8$ found above,
\begin{align}\label{eq:Acond}
A(123\{45\})
=\alpha_{12345}
-\frac{s_{23}}{2s_{12}}\gamma_{12}-\frac{s_{12}}{2s_{23}}\gamma_{23}
+\frac{s_{45}}{2}\left(\frac{\gamma_{12}}{s_{12}}+\frac{\gamma_{23}}{s_{23}}-\frac{\gamma_{45}}{s_{45}}\right),
\end{align}
which can be used to generate a consistency condition on $\alpha_{12345}$.
One cycles
\begin{align}\label{eq:leg5cycle}
A(123\{45\})&=A(12345)+A(12354),\nonumber\\
A(412\{35\})&=A(12354)+A(12534),\nonumber\\
A(341\{25\})&=A(12534)+A(15234),\nonumber\\
A(234\{15\})&=A(15234)+A(12345),
\end{align}
where we have used the cyclic condition from eq. (\ref{eq:Asymmetry})
to pull the $p_5$ leg around $A$.\footnote{The
	same technique was used in \cite{Bjerrum-Bohr:2013iza} to explore loop-momentum dependence of colour-dual $n$-gons in $\mathcal{N}=4$ SYM.  The cyclic nature of (\ref{eq:Xsymmetries}) suggests a similar $n$-gon structure in $X$.}
As a result,
\begin{align}
0&=A(123\{45\})-A(412\{35\})+A(341\{25\})-A(234\{15\})\nonumber\\
&=\alpha_{12345}-\alpha_{41235}+\alpha_{34125}-\alpha_{23415}\nonumber\\
&\qquad-\gamma_{25}-\gamma_{45}
-\frac{1}{2}\left(s_{15}+s_{35}\right)
\left(\frac{\gamma_{12}}{s_{12}}+\frac{\gamma_{23}}{s_{23}}+\frac{\gamma_{34}}{s_{34}}+\frac{\gamma_{41}}{s_{14}}\right).
\end{align}
Because $\beta_{12345}$ is not a valid solution for $\alpha_{12345}$, this is inconsistent.
A solution with minimal power counting in loop momentum is therefore not possible.

One might question whether a solution with minimal power counting is possible
if the three symmetry conditions on $X$ in eq. (\ref{eq:Xsymmetries}) are relaxed.
In this case, an intriguing solution to the box-triangle cut is
\begin{align}\label{eq:oldsoln}
X(12345;\ell_1,\ell_2)=\frac{1}{2s_{45}}n^{[\mathcal{N}=4]}\bigg(\usegraph{9}{delta431l}\!\bigg),
\end{align}
where we have used the fact that
\begin{align}
n^{[\mathcal{N}=4]}\bigg(\usegraph{9}{delta431l}\!\bigg)
=s_{45}\,\beta_{12345}+\gamma_{45}(\ell_1+p_5)^2
\end{align}
when $\ell_1^2=(\ell_1-p_1)^2=(\ell_1-p_{12})^2=(\ell_1+p_{45})^2=0$.\footnote{A
	similar structure appears in the double-box numerator (\ref{eq:n331}): the $\mathcal{N}=4$ numerator cancels away the nonlocal factor $s^{-1}$.}
However, this solution, or any other adhering to the form (\ref{eq:minansatz}),
fails to reproduce an off-shell symmetry of the nonplanar graph introduced in eq. (\ref{eq:n332}):
\begin{align}
n\bigg(\usegraph{9}{delta332l}\bigg)=-n\bigg(\usegraph{9}{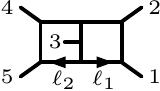}\bigg).
\end{align}
As explained in section \ref{sec:5ptsymmetry},
the symmetries (\ref{eq:Xsymmetries}) automatically imply this condition.

\subsection{Solving the ansatz}

Given the failure of a minimal solution,
we now explore a more general class of solutions depending on higher powers of loop momenta.
Still requiring that $X$ should be local in loop momenta,
we generalise eq. (\ref{eq:minansatz}) to a degree $\text{deg}(X)$ polynomial of the scalar products $z_i$:
\begin{align}\label{eq:Xansatz}
X(12345;\ell_1,\ell_2)
=\sum_{k=0}^{\text{deg}(X)}\sum_{1\leqslant i_1\leqslant\ldots\leqslant i_k\leqslant11}
A_{i_1\ldots i_k}(12345)z_{i_1}\ldots z_{i_k},
\end{align}
where the loop-momentum-independent functions $A_{i_1\cdots i_k}$
are the coefficients of the linearly independent monomials $z_{i_1}\ldots z_{i_k}$,
carrying total degree $k$.
For a given value of $k$,
we require a total of $11(11+1)\ldots(11+k)/k!$ functions $A_{i_1\ldots i_k}$ to account for all
possible monomials.
Loop-momentum dependence in our solution is now explicit:
it remains to identify dependence on the external momenta
which lives inside the functions $A_{i_1\ldots i_k}$.

Following the procedure used in the previous section,
we make considerable progress by solving two-term functional identities for
$A_{i_1\ldots i_k}$ that come from the symmetries in eq. (\ref{eq:Xsymmetries}).
Again, the relabelling property ensures the validity of these for any ordering of external momenta.
It is also possible to solve some identities coming from the
two cuts eqs. (\ref{eq:cut2205BFZ}) and (\ref{eq:cut2205LBFZ}).
For the former, the cut conditions are $z_1=z_4=z_5=z_7=0$;
for the latter, $z_1=z_3=z_5=z_7=0$.
However, these techniques are not sufficient to provide a unique solution for $X$,
so we make an ansatz for the remaining independent functions.

In view of our discussion in section~\ref{sec:nonlocal},
we know that the functions $A_{i_1\cdots i_k}$ may be nonlocal in the kinematic invariants.
We therefore define bases of linearly independent nonlocal monomials,
with degree of nonlocality $n$ and dimension $1+2m-2n$, as
\begin{align}\label{eq:nlbasis}
\Gamma^{[m,n]}=\left\{\left.\gamma_{ij}\frac{\prod_{k=1}^{m}s_k}{\prod_{l=1}^ns_l}\right|s_k\in\left\{s_{ij}\right\}\right\},
\end{align}
carrying linear dependence on $\gamma_{ij}$.
The lengths of these bases for various values of $m$ and $n$ are summarised in table \ref{tab:nlbases}.
Linear relations between such objects are generated by the identity
\begin{align}\label{eq:gammaidentity}
0=(\gamma_{12}+\gamma_{13})(s_{23}-s_{45})+\gamma_{23}(s_{12}-s_{13})+\gamma_{45}(s_{14}-s_{15}),
\end{align}
as well as the ability to cancel $s_{ij}$ terms between numerators and denominators.

%%%%%%%%%% TABLE %%%%%%%%%%
\begin{table*}
\centering
\begin{tabular}{| c | c c c c |}
\hline
& $m=0$ & $m=1$ & $m=2$ & $m=3$ \\
\hline
$n=0$ & 6 	& 25 	& 66 	& 140 \\
$n=1$ & 60  & 196 	& 435 	& 806 \\
$n=2$ & 300 & 790 	& 1536 	& 2585 \\
$n=3$ & 966	& 2185	& 3885	& 6131 \\
\hline
\end{tabular}
\caption{\small
The number of linearly independent basis elements of the nonlocal $\gamma$-bases $\Gamma^{[m,n]}$.
When $n=0$, there are $\frac{1}{12}m^4+\frac{4}{3}m^3+\frac{77}{12}m^2+\frac{67}{6}m+6$ elements.}
\label{tab:nlbases}
\end{table*}
%%%%%%%%%%%%%%%%%%%%%%%%%%%

These bases of nonlocal monomials allow us to make nonlocal ans\"{a}tze,
with degree of nonlocality $n$, for $A_{i_1\ldots i_k}$:
\begin{align}\label{eq:gammaansatz}
A_{i_1\ldots i_k}(12345)&=\sum_ja_{i_1\ldots i_k;j}\,\Gamma_j^{[n-k,n]},
\end{align}
where $a_{i_1\ldots i_k;j}$ are rational numbers and $j$ spans the length of the nonlocal basis.
Of course,
$A_{i_1\ldots i_k}$ applied to other orderings of the external legs
generates nonlocal monomials outside of the bases prescribed above.
It is necessary to rewrite these monomials in terms of the linearly independent basis elements
using identities such as the one in eq. (\ref{eq:gammaidentity}).

The procedure for solving for $X$ is now clear.
Having eliminated as many functions $A_{i_1\ldots i_k}$ as possible using functional identities,
one directly applies the ansatz decompositions given in eq. (\ref{eq:gammaansatz}) to the remaining functions.
Identifying coefficients of these bases leaves a large system of linear equations
for the rational numbers $a_{i_1\ldots i_k;j}$.
We find that a valid solution requires $\text{deg}(X)\geqslant3$:
this implies up to 6 powers of loop momentum in $X$,
so a total of 12 powers in the final expression for the pentabox numerator.
An initial ansatz with 10,004 parameters left 1260 undetermined once all conditions were accounted for.

%%%%%%%%%% TABLE %%%%%%%%%%
\begin{table*}
\centering
\begin{tabular}{| c | c | c |}
\hline
Function    & Known Symmetries                                      & Degrees of Freedom \\
\hline
$A$         & $A(12345)=A(23451)=-A(54321)$                         & 139 \\
$A_1$       & $A_1(12345)=-A_1(54321)$                              & 139 \\
$A_{1,2}$   & $A_{1,2}(12345)=-A_{1,2}(15432)$                      & 139 \\
$A_{1,3}$   & $A_{1,3}(12345)=-A_{1,3}(21543)$                      & 139 \\
$A_{1,5}$   & $A_{1,5}(12345)=-A_{1,5}(54321)$                      & 215 \\
$A_{1,6}$   & $A_{1,6}(12345)=-A_{1,6}(43215)$                      & 139 \\
$A_{1,7}$   & $A_{1,7}(12345)=-A_{1,7}(32154)$                      & 139 \\
$A_{1,1,6}$ & $A_{1,1,6}(12345)=A_{1,1,6}(52341)$                   & 139 \\
$A_{1,2,5}$ &                                                       & 214 \\
$A_{1,2,6}$ &                                                       & 139 \\
$A_{1,2,7}$ & $A_{1,2,7}(12345)=-A_{1,2,7}(15432)=A_{1,2,7}(12435)$ & 139 \\
$A_{1,3,5}$ &                                                       & 139 \\
$A_{1,3,6}$ &                                                       & 215 \\
\hline
\end{tabular}
\caption{\small
  Remaining degrees of freedom and known symmetries in the 13 independent functions specifying our solution for $X$.  Complete expressions with the extra degrees of freedom set to zero may be found in the attached ancillary file.}
\label{tab:Adof}
\end{table*}
%%%%%%%%%%%%%%%%%%%%%%%%%%%

We then further constrained the system by setting $A_{1,3,3}=A_{1,3,10}=A_{1,7,7}=0$,
as well as extracting an overall factor of $s_{12}^{-1}$ in $A_{1,3}(12345)$:
this left 215 undetermined parameters.
The final solution is completely specified by 13 nonzero functions $A_{i_1\ldots i_k}$,
given with their known symmetry properties and remaining degrees of freedom in table \ref{tab:Adof}.
In the attached ancillary file we provide the full solution for $X$,
having set these remaining degrees of freedom set to zero.

\section{Conclusions}\label{sec:conclusions}

The two-loop five-point amplitude discussed in this paper is a relatively simple object. Nevertheless, as we have shown, this amplitude is sufficiently complex that the problem of constructing a set of BCJ numerators is non-trivial. The numerators we found contain more powers of loop momentum than one would expect on the basis of the Feynman rules. They also contain spurious singularities in kinematic invariants.

Of course, it is always possible to add more loop-momentum dependence and additional spurious singularities to numerators of diagrams. The physical requirement is that any cut must take on its physical value. There is a large space of numerators which satisfy this condition. We exploited the freedom of adding more terms in order to build a numerator which satisfies the requirements of colour-kinematics duality. Given the generous freedom available, it is tempting to speculate that one can always find colour-dual numerators this way.

In this article, we chose to deal with the obstructions we encountered to the existence of colour-dual numerators with more traditional power counting by adding higher powers of loop momenta. Introducing large amounts of loop momenta in numerators at first seems like a bad idea, because the size of numerator ans\"atze grow quickly as more loop momentum dependence is allowed. We maintained control of our ansatz by imposing a powerful symmetry on the unknown function $X$. This symmetry allowed us to consider numerators with much more loop momentum than is typically possible. Finding a deeper understanding of the importance of this symmetry may help to find similar symmetry requirements in other cases. Recently, another route toward finding valid presentations of an amplitude which enjoy some double copy relationship with gravity was described~\cite{Bern:2015ooa}. The idea is that one can choose to impose the Jacobi identities only on a spanning set of cuts. This approach has the advantage that large amounts of extra loop momenta are not needed, which could be particularly advantageous in the context of studying UV divergences of the double copied amplitude. 

An interesting aspect of our work is that a connection between all-plus amplitudes in pure Yang-Mills theory and MHV amplitudes in $\mathcal N = 4$ SYM survives at two loops, even in the context of colour-kinematics duality. A family of colour-dual MHV numerators in the maximally supersymmetric theory at one loop was recently discovered~\cite{He:2015wgf}, built on our understanding of colour-kinematics duality in the self-dual theory~\cite{Monteiro:2011pc, Boels:2013bi}. 
Even though two-loop all-plus amplitudes are not self dual, it may be that insight into the nature of the residual link at two loops will allow for progress on one side to be recycled into the other.

%%%%%%%%%%%%%%%%%%%%%%%%%%%%%%%%%%%%%%%%%%%%%%%%%%%%
\begin{acknowledgments}
 We would like to thank Simon Badger, John Joseph Carrasco, Tristan Dennen, Einan Gardi, Henrik Johansson, Isobel Nicholson, Alexander Ochirov, Robbert Rietkerk and Radu Roiban for helpful discussions.  GM is supported by an STFC Studentship ST/K501980/1. DOC is supported in part by the STFC consolidated grant ``Particle Physics at the Higgs Centre'', by the National Science Foundation under grant NSF PHY11-25915, and by the Marie Curie FP7 grant 631370.
\end{acknowledgments}
%%%%%%%%%%%%%%%%%%%%%%%%%%%%%%%%%%%%%%%%%%%%%%%%%%%%

%%%%%%%%%%%%%%%%%%%%%%%%%%%%%%%%%%%%%%%%%%%%%%%%%%%%
\bibliographystyle{JHEP}
\bibliography{allplusbcj}

\providecommand{\href}[2]{#2}\begingroup\raggedright\begin{thebibliography}{10}

\bibitem{Roiban:2010kk}
R.~Roiban, {\it {Review of AdS/CFT Integrability, Chapter V.1: Scattering
  Amplitudes - a Brief Introduction}},  {\em Lett. Math. Phys.} {\bf 99} (2012)
  455--479, [\href{http://arxiv.org/abs/1012.4001}{{\tt arXiv:1012.4001}}].

\bibitem{Ellis:2011cr}
R.~Ellis, Z.~Kunszt, K.~Melnikov, and G.~Zanderighi, {\it {One-loop
  calculations in quantum field theory: from Feynman diagrams to unitarity
  cuts}},  \href{http://arxiv.org/abs/1105.4319}{{\tt arXiv:1105.4319}}.

\bibitem{Bern:2011qt}
Z.~Bern and Y.-t. Huang, {\it {Basics of Generalized Unitarity}},  {\em J.
  Phys.} {\bf A44} (2011) 454003, [\href{http://arxiv.org/abs/1103.1869}{{\tt
  arXiv:1103.1869}}].

\bibitem{Carrasco:2011hw}
J.~J.~M. Carrasco and H.~Johansson, {\it {Generic multiloop methods and
  application to N=4 super-Yang-Mills}},  {\em J. Phys.} {\bf A44} (2011)
  454004, [\href{http://arxiv.org/abs/1103.3298}{{\tt arXiv:1103.3298}}].

\bibitem{Dixon:2013uaa}
L.~J. Dixon, {\it {A brief introduction to modern amplitude methods}},  in {\em
  {Proceedings, 2012 European School of High-Energy Physics (ESHEP 2012)}},
  pp.~31--67, 2014.
\newblock \href{http://arxiv.org/abs/1310.5353}{{\tt arXiv:1310.5353}}.

\bibitem{Elvang:2013cua}
H.~Elvang and Y.-t. Huang, {\it {Scattering Amplitudes}},
  \href{http://arxiv.org/abs/1308.1697}{{\tt arXiv:1308.1697}}.

\bibitem{Carrasco:2015iwa}
J.~J.~M. Carrasco, {\it {Gauge and Gravity Amplitude Relations}},
  \href{http://arxiv.org/abs/1506.00974}{{\tt arXiv:1506.00974}}.

\bibitem{Bern:2008qj}
Z.~Bern, J.~Carrasco, and H.~Johansson, {\it {New Relations for Gauge-Theory
  Amplitudes}},  {\em Phys.Rev.} {\bf D78} (2008) 085011,
  [\href{http://arxiv.org/abs/0805.3993}{{\tt arXiv:0805.3993}}].

\bibitem{Bern:2010ue}
Z.~Bern, J.~J.~M. Carrasco, and H.~Johansson, {\it {Perturbative Quantum
  Gravity as a Double Copy of Gauge Theory}},  {\em Phys.Rev.Lett.} {\bf 105}
  (2010) 061602, [\href{http://arxiv.org/abs/1004.0476}{{\tt
  arXiv:1004.0476}}].

\bibitem{Bern:2010yg}
Z.~Bern, T.~Dennen, Y.-t. Huang, and M.~Kiermaier, {\it {Gravity as the Square
  of Gauge Theory}},  {\em Phys.Rev.} {\bf D82} (2010) 065003,
  [\href{http://arxiv.org/abs/1004.0693}{{\tt arXiv:1004.0693}}].

\bibitem{BjerrumBohr:2009rd}
N.~Bjerrum-Bohr, P.~H. Damgaard, and P.~Vanhove, {\it {Minimal Basis for Gauge
  Theory Amplitudes}},  {\em Phys.Rev.Lett.} {\bf 103} (2009) 161602,
  [\href{http://arxiv.org/abs/0907.1425}{{\tt arXiv:0907.1425}}].

\bibitem{Stieberger:2009hq}
S.~Stieberger, {\it {Open \& Closed vs. Pure Open String Disk Amplitudes}},
  \href{http://arxiv.org/abs/0907.2211}{{\tt arXiv:0907.2211}}.

\bibitem{Feng:2010my}
B.~Feng, R.~Huang, and Y.~Jia, {\it {Gauge Amplitude Identities by On-shell
  Recursion Relation in S-matrix Program}},  {\em Phys.Lett.} {\bf B695} (2011)
  350--353, [\href{http://arxiv.org/abs/1004.3417}{{\tt arXiv:1004.3417}}].

\bibitem{BjerrumBohr:2012mg}
N.~E.~J. Bjerrum-Bohr, P.~H. Damgaard, R.~Monteiro, and D.~O'Connell, {\it
  {Algebras for Amplitudes}},  {\em JHEP} {\bf 06} (2012) 061,
  [\href{http://arxiv.org/abs/1203.0944}{{\tt arXiv:1203.0944}}].

\bibitem{Cachazo:2012uq}
F.~Cachazo, {\it {Fundamental BCJ Relation in N=4 SYM From The Connected
  Formulation}},  \href{http://arxiv.org/abs/1206.5970}{{\tt arXiv:1206.5970}}.

\bibitem{Mafra:2011kj}
C.~R. Mafra, O.~Schlotterer, and S.~Stieberger, {\it {Explicit BCJ Numerators
  from Pure Spinors}},  {\em JHEP} {\bf 1107} (2011) 092,
  [\href{http://arxiv.org/abs/1104.5224}{{\tt arXiv:1104.5224}}].

\bibitem{Kawai:1985xq}
H.~Kawai, D.~Lewellen, and S.~Tye, {\it {A Relation Between Tree Amplitudes of
  Closed and Open Strings}},  {\em Nucl.Phys.} {\bf B269} (1986) 1.

\bibitem{Boels:2013bi}
R.~H. Boels, R.~S. Isermann, R.~Monteiro, and D.~O'Connell, {\it
  {Colour-Kinematics Duality for One-Loop Rational Amplitudes}},  {\em JHEP}
  {\bf 1304} (2013) 107, [\href{http://arxiv.org/abs/1301.4165}{{\tt
  arXiv:1301.4165}}].

\bibitem{He:2015wgf}
S.~He, R.~Monteiro, and O.~Schlotterer, {\it {String-inspired BCJ numerators
  for one-loop MHV amplitudes}},  \href{http://arxiv.org/abs/1507.06288}{{\tt
  arXiv:1507.06288}}.

\bibitem{Bern:2012uf}
Z.~Bern, J.~Carrasco, L.~Dixon, H.~Johansson, and R.~Roiban, {\it {Simplifying
  Multiloop Integrands and Ultraviolet Divergences of Gauge Theory and Gravity
  Amplitudes}},  {\em Phys.Rev.} {\bf D85} (2012) 105014,
  [\href{http://arxiv.org/abs/1201.5366}{{\tt arXiv:1201.5366}}].

\bibitem{Bern:2011rj}
Z.~Bern, C.~Boucher-Veronneau, and H.~Johansson, {\it {N $\ge$ 4 Supergravity
  Amplitudes from Gauge Theory at One Loop}},  {\em Phys.Rev.} {\bf D84} (2011)
  105035, [\href{http://arxiv.org/abs/1107.1935}{{\tt arXiv:1107.1935}}].

\bibitem{BoucherVeronneau:2011qv}
C.~Boucher-Veronneau and L.~Dixon, {\it {N $\ge$ 4 Supergravity Amplitudes from
  Gauge Theory at Two Loops}},  {\em JHEP} {\bf 1112} (2011) 046,
  [\href{http://arxiv.org/abs/1110.1132}{{\tt arXiv:1110.1132}}].

\bibitem{Bern:2012cd}
Z.~Bern, S.~Davies, T.~Dennen, and Y.-t. Huang, {\it {Absence of Three-Loop
  Four-Point Divergences in N=4 Supergravity}},  {\em Phys.Rev.Lett.} {\bf 108}
  (2012) 201301, [\href{http://arxiv.org/abs/1202.3423}{{\tt
  arXiv:1202.3423}}].

\bibitem{Bern:2012gh}
Z.~Bern, S.~Davies, T.~Dennen, and Y.-t. Huang, {\it {Ultraviolet Cancellations
  in Half-Maximal Supergravity as a Consequence of the Double-Copy Structure}},
   {\em Phys.Rev.} {\bf D86} (2012) 105014,
  [\href{http://arxiv.org/abs/1209.2472}{{\tt arXiv:1209.2472}}].

\bibitem{Carrasco:2012ca}
J.~J.~M. Carrasco, M.~Chiodaroli, M.~Gunaydin, and R.~Roiban, {\it {One-loop
  four-point amplitudes in pure and matter-coupled ${\cal N} \le 4$
  supergravity}},  {\em JHEP} {\bf 1303} (2013) 056,
  [\href{http://arxiv.org/abs/1212.1146}{{\tt arXiv:1212.1146}}].

\bibitem{Bern:2013yya}
Z.~Bern, S.~Davies, T.~Dennen, Y.-t. Huang, and J.~Nohle, {\it
  {Color-Kinematics Duality for Pure Yang-Mills and Gravity at One and Two
  Loops}},  \href{http://arxiv.org/abs/1303.6605}{{\tt arXiv:1303.6605}}.

\bibitem{Nohle:2013bfa}
J.~Nohle, {\it {Color-Kinematics Duality in One-Loop Four-Gluon Amplitudes with
  Matter}},  \href{http://arxiv.org/abs/1309.7416}{{\tt arXiv:1309.7416}}.

\bibitem{Bern:2013qca}
Z.~Bern, S.~Davies, and T.~Dennen, {\it {The Ultraviolet Structure of
  Half-Maximal Supergravity with Matter Multiplets at Two and Three Loops}},
  {\em Phys.Rev.} {\bf D88} (2013) 065007,
  [\href{http://arxiv.org/abs/1305.4876}{{\tt arXiv:1305.4876}}].

\bibitem{Bern:2013uka}
Z.~Bern, S.~Davies, T.~Dennen, A.~V. Smirnov, and V.~A. Smirnov, {\it {The
  Ultraviolet Properties of N=4 Supergravity at Four Loops}},
  \href{http://arxiv.org/abs/1309.2498}{{\tt arXiv:1309.2498}}.

\bibitem{Carrasco:2013ypa}
J.~Carrasco, R.~Kallosh, R.~Roiban, and A.~Tseytlin, {\it {On the U(1) duality
  anomaly and the S-matrix of N=4 supergravity}},  {\em JHEP} {\bf 1307} (2013)
  029, [\href{http://arxiv.org/abs/1303.6219}{{\tt arXiv:1303.6219}}].

\bibitem{Bern:2014lha}
Z.~Bern, S.~Davies, and T.~Dennen, {\it {The Ultraviolet Critical Dimension of
  Half-Maximal Supergravity at Three Loops}},
  \href{http://arxiv.org/abs/1412.2441}{{\tt arXiv:1412.2441}}.

\bibitem{Bern:2014sna}
Z.~Bern, S.~Davies, and T.~Dennen, {\it {Enhanced ultraviolet cancellations in
  $\mathcal N=5$ supergravity at four loops}},  {\em Phys. Rev.} {\bf D90}
  (2014), no.~10 105011, [\href{http://arxiv.org/abs/1409.3089}{{\tt
  arXiv:1409.3089}}].

\bibitem{Bern:2015xsa}
Z.~Bern, C.~Cheung, H.-H. Chi, S.~Davies, L.~Dixon, and J.~Nohle, {\it
  {Evanescent Effects Can Alter Ultraviolet Divergences in Quantum Gravity
  without Physical Consequences}},  \href{http://arxiv.org/abs/1507.06118}{{\tt
  arXiv:1507.06118}}.

\bibitem{Broedel:2012rc}
J.~Broedel and L.~J. Dixon, {\it {Color-kinematics duality and double-copy
  construction for amplitudes from higher-dimension operators}},  {\em JHEP}
  {\bf 10} (2012) 091, [\href{http://arxiv.org/abs/1208.0876}{{\tt
  arXiv:1208.0876}}].

\bibitem{Chiodaroli:2013upa}
M.~Chiodaroli, Q.~Jin, and R.~Roiban, {\it {Color/kinematics duality for
  general abelian orbifolds of N=4 super Yang-Mills theory}},  {\em JHEP} {\bf
  1401} (2014) 152, [\href{http://arxiv.org/abs/1311.3600}{{\tt
  arXiv:1311.3600}}].

\bibitem{Chiodaroli:2014xia}
M.~Chiodaroli, M.~Gunaydin, H.~Johansson, and R.~Roiban, {\it {Scattering
  amplitudes in N=2 Maxwell-Einstein and Yang-Mills/Einstein supergravity}},
  \href{http://arxiv.org/abs/1408.0764}{{\tt arXiv:1408.0764}}.

\bibitem{Johansson:2014zca}
H.~Johansson and A.~Ochirov, {\it {Pure Gravities via Color-Kinematics Duality
  for Fundamental Matter}},  \href{http://arxiv.org/abs/1407.4772}{{\tt
  arXiv:1407.4772}}.

\bibitem{Johansson:2015oia}
H.~Johansson and A.~Ochirov, {\it {Color-Kinematics Duality for QCD
  Amplitudes}},  \href{http://arxiv.org/abs/1507.00332}{{\tt
  arXiv:1507.00332}}.

\bibitem{Chiodaroli:2015rdg}
M.~Chiodaroli, M.~Gunaydin, H.~Johansson, and R.~Roiban, {\it {Spontaneously
  Broken Yang-Mills-Einstein Supergravities as Double Copies}},
  \href{http://arxiv.org/abs/1511.01740}{{\tt arXiv:1511.01740}}.

\bibitem{Berkovits:2000fe}
N.~Berkovits, {\it {Super Poincare covariant quantization of the superstring}},
   {\em JHEP} {\bf 04} (2000) 018,
  [\href{http://arxiv.org/abs/hep-th/0001035}{{\tt hep-th/0001035}}].

\bibitem{Mafra:2014oia}
C.~R. Mafra and O.~Schlotterer, {\it {Multiparticle SYM equations of motion and
  pure spinor BRST blocks}},  \href{http://arxiv.org/abs/1404.4986}{{\tt
  arXiv:1404.4986}}.

\bibitem{Mafra:2014gja}
C.~R. Mafra and O.~Schlotterer, {\it {Towards one-loop SYM amplitudes from the
  pure spinor BRST cohomology}},  {\em Fortsch. Phys.} {\bf 63} (2015), no.~2
  105--131, [\href{http://arxiv.org/abs/1410.0668}{{\tt arXiv:1410.0668}}].

\bibitem{Mafra:2015mja}
C.~R. Mafra and O.~Schlotterer, {\it {Two-loop five-point amplitudes of super
  Yang-Mills and supergravity in pure spinor superspace}},
  \href{http://arxiv.org/abs/1505.02746}{{\tt arXiv:1505.02746}}.

\bibitem{Bargheer:2012gv}
T.~Bargheer, S.~He, and T.~McLoughlin, {\it {New Relations for
  Three-Dimensional Supersymmetric Scattering Amplitudes}},  {\em
  Phys.Rev.Lett.} {\bf 108} (2012) 231601,
  [\href{http://arxiv.org/abs/1203.0562}{{\tt arXiv:1203.0562}}].

\bibitem{Huang:2012wr}
Y.-t. Huang and H.~Johansson, {\it {Equivalent D=3 Supergravity Amplitudes from
  Double Copies of Three-Algebra and Two-Algebra Gauge Theories}},  {\em
  Phys.Rev.Lett.} {\bf 110} (2013) 171601,
  [\href{http://arxiv.org/abs/1210.2255}{{\tt arXiv:1210.2255}}].

\bibitem{Huang:2013kca}
Y.-t. Huang, H.~Johansson, and S.~Lee, {\it {On Three-Algebra and
  Bi-Fundamental Matter Amplitudes and Integrability of Supergravity}},  {\em
  JHEP} {\bf 1311} (2013) 050, [\href{http://arxiv.org/abs/1307.2222}{{\tt
  arXiv:1307.2222}}].

\bibitem{Boels:2012ew}
R.~H. Boels, B.~A. Kniehl, O.~V. Tarasov, and G.~Yang, {\it {Color-kinematic
  Duality for Form Factors}},  {\em JHEP} {\bf 1302} (2013) 063,
  [\href{http://arxiv.org/abs/1211.7028}{{\tt arXiv:1211.7028}}].

\bibitem{Bern:2011ia}
Z.~Bern and T.~Dennen, {\it {A Color Dual Form for Gauge-Theory Amplitudes}},
  {\em Phys. Rev. Lett.} {\bf 107} (2011) 081601,
  [\href{http://arxiv.org/abs/1103.0312}{{\tt arXiv:1103.0312}}].

\bibitem{Monteiro:2011pc}
R.~Monteiro and D.~O'Connell, {\it {The Kinematic Algebra From the Self-Dual
  Sector}},  {\em JHEP} {\bf 1107} (2011) 007,
  [\href{http://arxiv.org/abs/1105.2565}{{\tt arXiv:1105.2565}}].

\bibitem{Saotome:2012vy}
R.~Saotome and R.~Akhoury, {\it {Relationship Between Gravity and Gauge
  Scattering in the High Energy Limit}},  {\em JHEP} {\bf 01} (2013) 123,
  [\href{http://arxiv.org/abs/1210.8111}{{\tt arXiv:1210.8111}}].

\bibitem{Monteiro:2014cda}
R.~Monteiro, D.~O'Connell, and C.~D. White, {\it {Black holes and the double
  copy}},  \href{http://arxiv.org/abs/1410.0239}{{\tt arXiv:1410.0239}}.

\bibitem{Luna:2015paa}
A.~Luna, R.~Monteiro, D.~O'Connell, and C.~D. White, {\it {The classical double
  copy for Taub–NUT spacetime}},  {\em Phys. Lett.} {\bf B750} (2015)
  272--277, [\href{http://arxiv.org/abs/1507.01869}{{\tt arXiv:1507.01869}}].

\bibitem{Bern:2011qn}
Z.~Bern, J.~J. Carrasco, L.~J. Dixon, H.~Johansson, and R.~Roiban, {\it
  {Amplitudes and Ultraviolet Behavior of N = 8 Supergravity}},  {\em Fortsch.
  Phys.} {\bf 59} (2011) 561--578, [\href{http://arxiv.org/abs/1103.1848}{{\tt
  arXiv:1103.1848}}].

\bibitem{Bern:2012uc}
Z.~Bern, J.~Carrasco, H.~Johansson, and R.~Roiban, {\it {The Five-Loop
  Four-Point Amplitude of N=4 super-Yang-Mills Theory}},  {\em Phys.Rev.Lett.}
  {\bf 109} (2012) 241602, [\href{http://arxiv.org/abs/1207.6666}{{\tt
  arXiv:1207.6666}}].

\bibitem{Bern:2015ooa}
Z.~Bern, S.~Davies, and J.~Nohle, {\it {Double-Copy Constructions and Unitarity
  Cuts}},  \href{http://arxiv.org/abs/1510.03448}{{\tt arXiv:1510.03448}}.

\bibitem{Badger:2013gxa}
S.~Badger, H.~Frellesvig, and Y.~Zhang, {\it {A Two-Loop Five-Gluon Helicity
  Amplitude in QCD}},  {\em JHEP} {\bf 1312} (2013) 045,
  [\href{http://arxiv.org/abs/1310.1051}{{\tt arXiv:1310.1051}}].

\bibitem{Badger:2015lda}
S.~Badger, G.~Mogull, A.~Ochirov, and D.~O’Connell, {\it {A Complete
  Two-Loop, Five-Gluon Helicity Amplitude in Yang-Mills Theory}},  {\em JHEP}
  {\bf 10} (2015) 064, [\href{http://arxiv.org/abs/1507.08797}{{\tt
  arXiv:1507.08797}}].

\bibitem{Gehrmann:2015bfy}
T.~Gehrmann, J.~M. Henn, and N.~A.~L. Presti, {\it {Analytic form of the
  two-loop planar five-gluon all-plus-helicity amplitude in QCD}},
  \href{http://arxiv.org/abs/1511.05409}{{\tt arXiv:1511.05409}}.

\bibitem{Carrasco:2011mn}
J.~J. Carrasco and H.~Johansson, {\it {Five-Point Amplitudes in N=4
  Super-Yang-Mills Theory and N=8 Supergravity}},  {\em Phys.Rev.} {\bf D85}
  (2012) 025006, [\href{http://arxiv.org/abs/1106.4711}{{\tt
  arXiv:1106.4711}}].

\bibitem{Bern:2002zk}
Z.~Bern, A.~De~Freitas, L.~J. Dixon, and H.~Wong, {\it {Supersymmetric
  regularization, two loop QCD amplitudes and coupling shifts}},  {\em
  Phys.Rev.} {\bf D66} (2002) 085002,
  [\href{http://arxiv.org/abs/hep-ph/0202271}{{\tt hep-ph/0202271}}].

\bibitem{tristanCode}
T.~Dennen. Personal communication.

\bibitem{Bern:2000dn}
Z.~Bern, L.~J. Dixon, and D.~Kosower, {\it {A Two loop four gluon helicity
  amplitude in QCD}},  {\em JHEP} {\bf 0001} (2000) 027,
  [\href{http://arxiv.org/abs/hep-ph/0001001}{{\tt hep-ph/0001001}}].

\bibitem{Badger:2012dp}
S.~Badger, H.~Frellesvig, and Y.~Zhang, {\it {Hepta-Cuts of Two-Loop Scattering
  Amplitudes}},  {\em JHEP} {\bf 1204} (2012) 055,
  [\href{http://arxiv.org/abs/1202.2019}{{\tt arXiv:1202.2019}}].

\bibitem{Bern:1996ja}
Z.~Bern, L.~J. Dixon, D.~C. Dunbar, and D.~A. Kosower, {\it {One loop selfdual
  and N=4 superYang-Mills}},  {\em Phys.Lett.} {\bf B394} (1997) 105--115,
  [\href{http://arxiv.org/abs/hep-th/9611127}{{\tt hep-th/9611127}}].

\bibitem{Broedel:2011pd}
J.~Broedel and J.~J.~M. Carrasco, {\it {Virtuous Trees at Five and Six Points
  for Yang-Mills and Gravity}},  {\em Phys. Rev.} {\bf D84} (2011) 085009,
  [\href{http://arxiv.org/abs/1107.4802}{{\tt arXiv:1107.4802}}].

\bibitem{Bjerrum-Bohr:2013iza}
N.~E.~J. Bjerrum-Bohr, T.~Dennen, R.~Monteiro, and D.~O'Connell, {\it
  {Integrand Oxidation and One-Loop Colour-Dual Numerators in N=4 Gauge
  Theory}},  {\em JHEP} {\bf 1307} (2013) 092,
  [\href{http://arxiv.org/abs/1303.2913}{{\tt arXiv:1303.2913}}].

\end{thebibliography}\endgroup
 
\end{document}